\documentclass[twocolumn,aps,pre,floats, showpacs]{revtex4}
\pdfoutput=1
\usepackage{graphics,graphicx,epsfig}
\usepackage{amssymb}
\usepackage{epsf,epstopdf,wrapfig}

\newcommand {\e}[1]{\mathrm{~#1}}    

\begin{document}

\title{Information capacity of genetic regulatory elements}

\author{Ga\v{s}per Tka\v{c}ik, Curtis G. Callan, Jr.$^*$ and William Bialek$^*$}

\affiliation{Joseph Henry Laboratories of Physics,  Lewis--Sigler Institute for Integrative Genomics, and $^*$Princeton Center for Theoretical Physics,
Princeton University, Princeton, New Jersey 08544}

\begin{abstract}
{Changes in a cell's external or internal conditions are usually reflected in the concentrations of the relevant transcription factors. These proteins in turn modulate the expression levels of the genes under their control and sometimes need to perform non-trivial computations that integrate several inputs and affect multiple genes. At the same time, the activities of the regulated genes would fluctuate even if the inputs were held fixed, as a consequence of the intrinsic noise in the system, and such noise must fundamentally limit the reliability of any genetic computation. Here we use information theory to formalize the notion of information transmission in simple genetic regulatory elements in the presence of physically realistic noise sources. The dependence of this ``channel capacity'' on noise parameters, cooperativity and cost of making signaling molecules is explored systematically. We find that, at least in principle, capacities higher than one bit should be achievable and that consequently genetic regulation is not limited the use of  binary, or ``on-off'', components.}
\end{abstract}

\date{\today}

\pacs{87.16.Yc, 87.16.Xa, 89.70.+c}

\maketitle

\section{Introduction}
Networks of interacting genes coordinate complex cellular processes, such as responding to stress, adapting the metabolism to a varying diet, maintaining the circadian cycle or producing an intricate spatial arrangement of differentiated cells during development \cite{gasch+al_00,leloup+goldbeter_03,lawrence_92,levine+davidson_05}. 
The success of such regulatory modules is at least partially characterized by their ability to produce reliable responses to repeated stimuli or changes in the environment over a wide dynamic range, and to perform the genetic computations reproducibly, either on a day-by-day or generation timescale. In doing so the regulatory elements are confronted by noise arising from physical processes that implement such genetic computations, and this noise ultimately traces its origins back to the fact that the state variables of the system are concentrations of chemicals and ``computations'' are really reactions between individual molecules, usually present at low copy numbers \cite{macadams+arkin_97,kepler+elston_01}.

It is useful to picture the regulatory module as a device that, given some input, computes an output, which in our case will be a set of expression levels of regulated genes. Sometimes the inputs to the module are easily identified, such as when they are the actual chemicals that a system detects and responds to, for example chemoattractant molecules, hormones or transcription factors (TFs). There are cases, however, when it is beneficial to think about the inputs on a more abstract level: in embryonic development we talk of ``positional information''  and think of the regulatory module as trying to produce a different gene expression footprint at each spatial location \cite{wolpert_69}; alternatively, circadian clocks generate distinguishable gene expression profiles corresponding to various phases of the day \cite{leloup+goldbeter_03}. Regardless of whether we view the input as a physical concentration of some transcription factor or perhaps a position within the embryo, and whether the computation is complicated or as simple as an inversion produced by a repressor, we want to quantify its reliability in the presence of noise, and ask what the biological system can do to maximize this reliability.

If we make many observations of a genetic regulatory element in its natural conditions we are collecting samples drawn from a distribution $p(\mathcal{I}, \mathcal{O})$, where $\mathcal{I}$ describes the state of the input and $\mathcal{O}$ the state of the output. Saying that the system is able to produce a reliable response $\mathcal{O}$ across the spectrum of naturally occurring input conditions $p(\mathcal{I})$ amounts to saying that the dependency -- either linear or strongly non-linear -- between the input and output is high, i.e. far from random. Shannon has shown how to associate a unique measure, the mutual information $I$, with the notion of dependency between two quantities drawn from a joint distribution \cite{shannon48,shannon49,cover+thomas}:

\begin{equation}
I(\mathcal{I}; \mathcal{O})=\int \!\!\! \int d\mathcal{I}\,d\mathcal{O}\,p(\mathcal{I}, \mathcal{O})\log_2\frac{p(\mathcal{I}, \mathcal{O})}{p(\mathcal{I})p(\mathcal{O})}. \label{genmutinf}
\end{equation}

The resulting quantity is a measure in bits and is essentially the logarithm of the number of states in the input that produce distinguishable outputs given the noise. A device that has one bit of capacity can be thought of as an ``on-off'' switch, two bits correspond to four distinguishable regulatory settings, and so on. Although the input is usually a continuous quantity, such as nutrient concentration or phase of the day, the noise present in the regulatory element corrupts the computation and does not allow the arbitrary resolution of a real-valued input to propagate to the output; instead, the mutual information tells us how precisely different inputs are distinguishable to the organism.

Experimental or theoretical characterization of the joint distribution, $p(\mathcal{I}, \mathcal{O})$, for a regulatory module can be very difficult if the inputs and outputs live in a high-dimensional space. We can proceed, nevertheless, by remembering that the building blocks of complex modules are much simpler, and finally must reduce to the point where a single gene is controlled by transcription factors that bind to its promoter region and  tune the level of its expression. While taking a simple element out of its network will not be illuminating about how the network as a whole behaves in general -- especially if there are feedback loops -- there may be cases where the information flow is ``bottlenecked'' through  a single gene, and its reliability will therefore limit that of the network. In addition, the analysis of  a simple regulatory element will provide directions for taking on more complicated systems; see Ref \cite{ziv+al_06} for a recent related analysis.

Our aim  is therefore to understand the reliability of a simple genetic regulatory element, that is, of  a single activator or repressor transcription factor controlling the expression level of its downstream gene. We will identify the concentration $c$ of the transcription factor as the only input, $\mathcal{I}\equiv \left\{ c\right\}$, and the expression level of the downstream gene $g$ as the only relevant output, $\mathcal{O}\equiv \left\{ g\right\}$. The regulatory element itself will be parametrized by input/output kernel, $p(g|c)$, i.e. the distribution (as opposed to a ``deterministic'' function $g=g(c)$ in case of a noiseless system) of possible outputs given that the input is fixed to some particular level $c$. For each such kernel, we will then compute the maximum amount of information, $I(c;g)$, that can be transmitted through it, and examine how this capacity depends on the properties of the kernel.
\begin{figure} 
   \centering
   \includegraphics[width=3in]{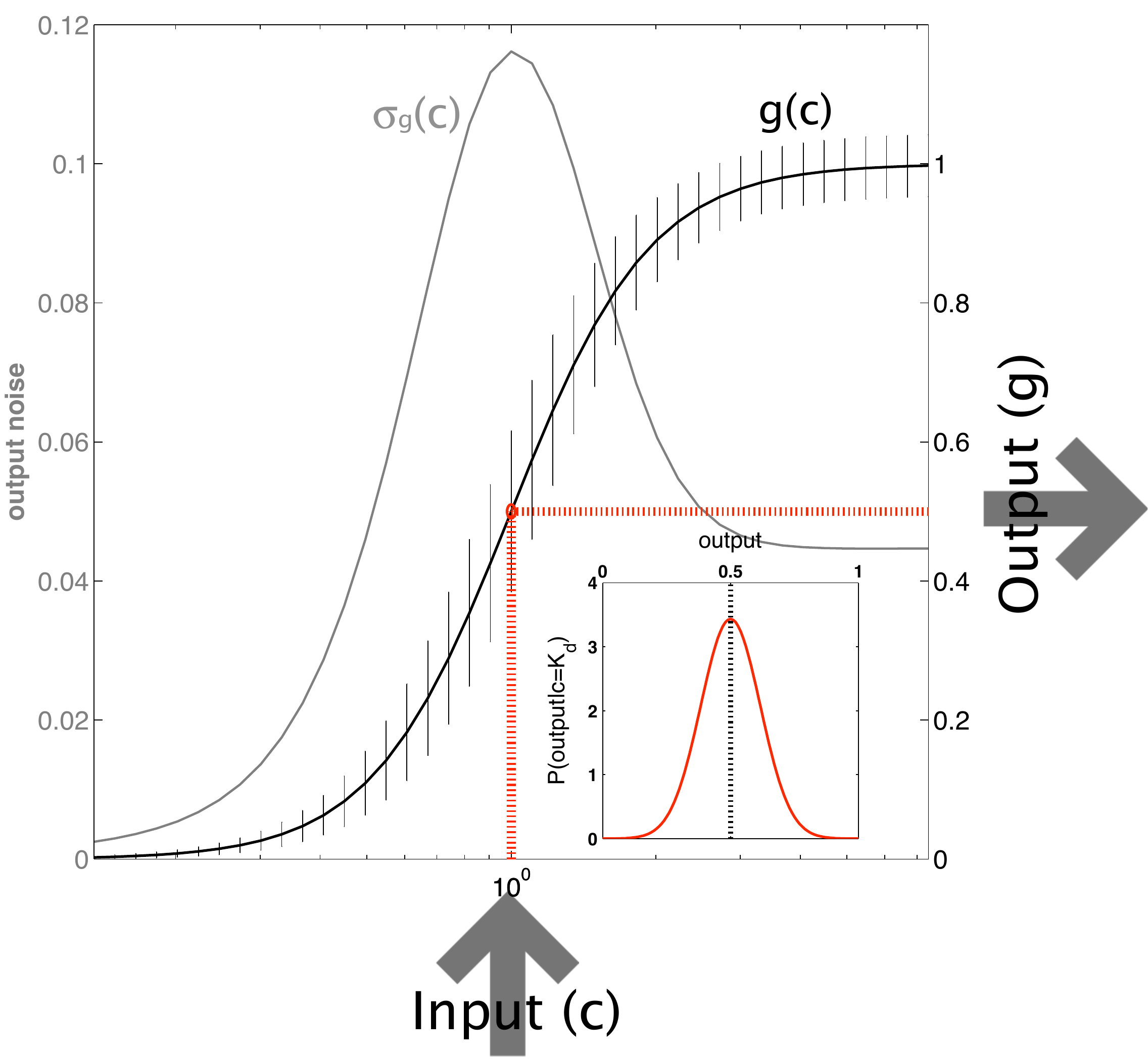} 
   \caption{(Color online) A schematic diagram of a simple regulatory element. Each input is mapped to a mean output according to the input/output relation (thick sigmoidal black line). Because the system is noisy, the output fluctuates about the mean. This noise is plotted in gray as a function of the input and shown in addition as error bars on the mean input/output relation. Inset shows the probability distribution of outputs at half saturation, $p(g|c=K_d)$ (red dotted lines); in this simple example we assume that the distribution is Gaussian and therefore fully characterized by its mean and variance.} 
   \label{f-schemeio2}
\end{figure}
\section{Maximizing information transmission}
The input/output kernel of a simple regulatory element, $p(g|c)$, is determined by the biophysics of transcription factor-DNA interaction, transcription and translation. In contrast, the distribution of inputs, $p(c)$, that the cell uses during its ``typical'' lifetime, is free for the cell to change. The cell's transcription factor expression footprint is its representation of the environment and internal state, and the form of this representation can be the target of adaptation or evolutionary processes. Together, the input/output kernel and the distribution of inputs define the joint distribution, $p(c,g)=p(g|c)\,p(c)$, and consequently the mutual information of Eq (\ref{genmutinf}) between  the input and the output, $I(c;g)$. 

Maximizing the information between the inputs and outputs, which corresponds to our notions of reliability in representation and computation, will therefore imply a specific matching between the given input/output kernel and the distribution of inputs, $p(c)$, that is being optimized. If one believes that a specific regulatory element has been tuned for maximal information transmission, then the optimal solution for the inputs, $p^*(c)$, and the resulting optimal distribution of outputs, $p^*(g)=\int dc\, p(g|c)p^*(c)$, become experimentally verifiable predictions. If, on the other hand, the system is not really maximizing information transmission, then the capacity achievable with a given kernel and its optimal input distribution, $I[p(g|c),p^*(c)]$, can still be regarded as a (hopefully revealing) upper bound on the true information transmission of the system. 

During the past decades the measurements of regulatory elements have focused on recovering the mean response of a gene under the control of a transcription factor that had its activity modulated by experimentally adjustable levels of inducer or inhibitor molecules \cite{mbg}. Typically, a sigmoidal response is observed with a single regulator, as in Fig \ref{f-schemeio2}, and more complicated regulatory ``surfaces'' are possible when there are two or more simultaneous inputs to the system \cite{setty+al_03,kuhlman+al_07}. In our notation, these experiments measure the conditional average over the distribution of outputs, $\bar{g}(c)=\int dg\, g\, p(g|c)$. 
Developments in flow cytometry and single-cell microscopy enabled the experimenters to start tracking in time and across the population of cells the expression levels of fluorescent reporter genes and thus open a window into the behavior of fluctuations. Consequently, work exploring the noise in gene expression, or $\sigma_g^2(c)=\int dg\, (g-\bar{g})^2 p(g|c)$, has begun to accumulate, on both the experimental and biophysical modeling side \cite{thattai+oudenaarden_01,elowitz+al_02,ozbudak+al_02,blake+al_03}. The efforts to further characterize and understand this noise were renewed by theoretical work by Swain and coworkers \cite{swain+al_02}
 that has shown how to separate intrinsic and extrinsic components of the noise, i.e. the noise due to the stochasticity of the observed regulatory process in a single cell, and the noise contribution that arises because typical experiments make many single-cell measurements and the internal chemical environments of these cells differ across the population.

\subsection{Small noise approximation}
 We start by showing how the optimal distributions can be computed analytically if the input/output kernel is Gaussian and the noise is small, and proceed by presenting the exact numerical solution later.
Let us assume then that the first and second moments of the conditional distribution are given, and write the input/output kernel as a set of Gaussian distributions $\mathcal{G}(g;\bar{g}(c),\sigma_g(c))$, or explicitly:
\begin{equation}
p(g|c) = {1\over{\sqrt{2\pi \sigma_g^2 (c)}}} \exp\Bigg{\{} - \, 
{{[g - \bar g (c)]^2}\over{2 \sigma_g^2 (c)}}  \Bigg{\}} , \label{gaussianioo}
\end{equation}
where both the mean response, $\bar{g}(c)$, and the noise, $\sigma_g(c)$, depend on the input, as illustrated in Fig \ref{f-schemeio2}.

We rewrite the mutual information between the input and the output of Eq (\ref{genmutinf}) in the following way:
\begin{eqnarray}
I(c;g)&=&\int dc\, p(c)\int dg\, p(g|c)\log_2 p(g|c)- \nonumber \\
&-&\int dc\, p(c)\int dg\, p(g|c) \log_2 p(g). \label{mutinf}
\end{eqnarray}
The first term can be evaluated exactly for Gaussian distributions, $p(g|c)=\mathcal{G}(g; \bar{g}(c),\sigma_g(c))$. The integral over $g$ is just the calculation of the (negative of the) entropy of the Gaussian, and the first term therefore evaluates to $-\langle S\left[\mathcal{G}(g;\bar{g},\sigma_g\right]  \rangle_{p(c)}=-\frac{1}{2}\langle \log_2{2\pi e \sigma^2_g(c)}\rangle_{p(c)}$.

In the second term of Eq (\ref{mutinf}), the integral over $g$ can be viewed as calculating $\langle \log_2 p(g)\rangle$ under the distribution $p(g|c)$. For an arbitrary continuous function $f(g)$ we can expand the integrals with the Gaussian measure around the mean:
\begin{eqnarray}
\langle f(g)\rangle_{\mathcal{G}(g;\bar{g},\sigma_g)}&=&\int dg\, \mathcal{G}(g)f(\bar{g})+ \nonumber \\
&+&\int dg\, \mathcal{G}(g)\frac{\partial f}{\partial g}\Big |_{\bar{g}}\left(g-\bar{g}\right)+ \nonumber \\
&+&\frac{1}{2}\int dg\, \mathcal{G}(g)\frac{\partial f^2}{\partial g^2}\Big |_{\bar{g}}\left(g-\bar{g}\right)^2+\cdots \label{mutexp}
\end{eqnarray}
The first term of the expansion simply evaluates to $f(\bar{g})$. The series expansion would end at the first term if we were to take the small noise limit, $\lim_{\sigma_g\rightarrow 0} \mathcal{G}(g;\bar{g},\sigma_g)=\delta(g-\bar{g})$. The second term of the expansion is zero because of symmetry, and the third term evaluates to $\frac{1}{2}\sigma_g^2 f''(\bar{g})$. We apply the expansion of Eq (\ref{mutexp}) and compute the second term in the expression for the mutual information, Eq (\ref{mutinf}), with $f(g)=\log_2 p(g)$. Taking only the zeroth order of the expansion, we get
\begin{equation}
I(c;g)=-\int dc\,p(c)\left[\log_2{\sqrt{2\pi e}\sigma_g(c)}+ \log_2 p(\bar{g}(c))\right];
\end{equation}
we can rewrite the probability distributions in terms of $\bar{g}$, using $p(c)\,dc=p(\bar{g})\,d\bar{g}$.  
To maximize the information transmission we form the following Lagrangian and introduce the multiplier $\Lambda$ that keeps the resulting distribution normalized:
\begin{equation}
\mathcal{L}[p(\bar{g})]=-\int d\bar{g}\,p(\bar{g})\log_2\left(\sqrt{2\pi e}\sigma_g(\bar{g}) p(\bar{g})\right)- \Lambda\int d\bar{g}\,p(\bar{g}).
\end{equation}
The optimal solution is obtained by taking a variational derivative with respect to $p(\bar{g})$, $\frac{\delta\mathcal{L}[p(\bar{g})]}{\delta p(\bar{g})}=0$. The solution is
\begin{equation}
p^*(\bar{g})=\frac{1}{Z}\cdot\frac{1}{\sigma_g(\bar{g})}. \label{optg}
\end{equation}
By inserting the optimal solution, Eq (\ref{optg}), into the expression for mutual information, Eq (\ref{mutinf}), we get an explicit result for the capacity:
\begin{equation}
I_{\rm opt} (c;g) = \log_2 \left[ {Z\over{\sqrt{2\pi e}}}\right] ,
\label{Iopt}
\end{equation}
where $Z$ is the normalization of the optimal solution in Eq (\ref{optg}):
\begin{equation}
Z=\int_0^1 \frac{d\bar{g}}{\sigma_g(\bar{g})}. \label{ZZ}
\end{equation}

The optimization with respect to the distribution of inputs, $p(c)$, has led us to the result for the optimal distribution of \emph{mean outputs}, Eq (\ref{optg}). We had to assume that the input/output kernel is Gaussian and that the noise is small, and we refer to this result as the small-noise approximation (SNA) for channel capacity. Note that in this approximation only the knowledge of the noise in the output as a function of mean output, $\sigma_g(\bar{g})$, matters for capacity computation and the direct dependence on the input $c$ is irrelevant. This is important because the behavior of intrinsic noise as a function of the mean output is an experimentally accessible quantity \cite{elowitz+al_02}. Note also that for big enough noise the normalization constant $Z$ will be small compared to $\sqrt{2\pi e}$, and the small-noise capacity approximation of Eq (\ref{Iopt}) will break down by predicting negative information values.
\subsection{Large noise approximation}
\begin{figure} 
   \centering
   \includegraphics[width=3in]{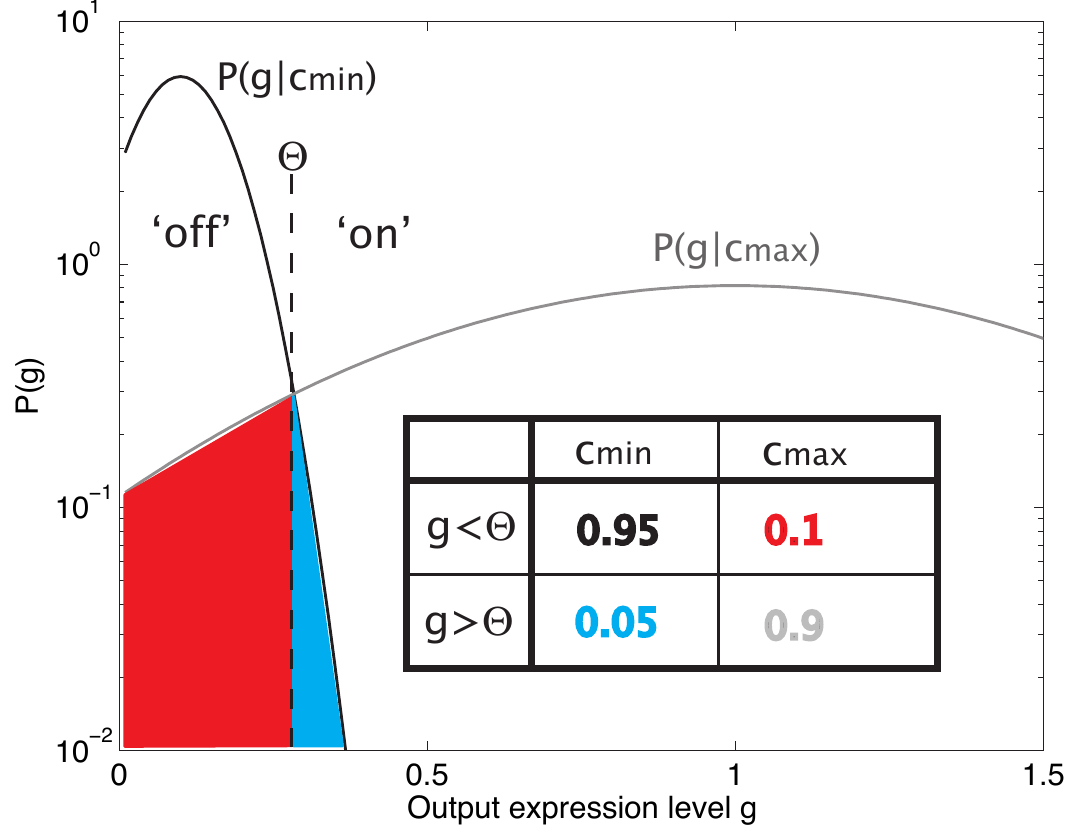} 
   \caption{(Color online) An illustration of the large noise approximation. We consider distributions of the output at minimal $(c_{\rm min})$ and full $(c_{\rm max})$ induction as trying to convey a single binary decision, and construct the corresponding encoding table (inset) by discretizing the output using the threshold $\Theta$. The capacity of such an asymmetric binary channel is degraded from the theoretical maximum of 1 bit, because the distributions overlap (blue and red). For unclipped Gaussians the optimal threshold $\Theta$ is at the intersection of two alternative pdfs, but in general one searches for the optimal $\Theta$ that maximizes information in Eq (\ref{bnas}).} 
   \label{f-bna}
   \end{figure}

Simple regulatory elements usually have a monotonic, saturating input/output relation, as shown in Fig \ref{f-schemeio2}, and (at least) a shot noise component whose variance scales with the mean. If the noise strength is increased, the information transmission must drop and, even with the optimally tuned input distribution, eventually yield only a bit or less of capacity. Intuitively, the best such a noisy system can do is to utilize only the lowest and highest achievable input concentrations, and ignore the continuous range in between. Thus, the mean responses will be as different as possible, and the noise at low expression will also be low because it scales with the mean. More formally, if only $\left\{c_{\rm min},c_{\rm max}\right\}$ are used as inputs, then the result is either $p(g|c_{\rm min})$ or $p(g|c_{\rm max})$; the optimization of channel capacity reduces to finding $p(c_{\rm min})$, with $p(c_{\rm max})=1-p(c_{\rm min})$. This problem can be solved by realizing that each of the two possible input concentrations produces their respective Gaussian output distributions, and by maximizing information by varying $p(c_{\rm min})$. Simplifying even further, we can threshold the outputs and allow $g$ to take on only two values instead of a continuous range; then, each of the two possible inputs, ``min'' and ``max'', maps into two possible outputs, ``on'' and ``off'', and confusion in the channel arises because ``min'' input might be misunderstood as ``on'' output and vice versa with probabilities given by the output distribution overlaps, as shown schematically in Fig \ref{f-bna}.

In the latter case we can use the analytic formula for the capacity of the binary asymmetric channel.
If $\eta$ is the probability of detecting an ``off'' output if ``max'' input was sent, and $\xi$ is a probability of receiving an ``off''  output if ``min'' input was sent, and $H(\cdot)$ is a binary entropy function:
\begin{equation}
H(p)=-p\log_2 p-(1-p)\log_2(1-p),
\end{equation}
then the capacity of such asymmetric channel is \cite{Silverman_55}:
\begin{equation}
I(c;g)=\frac{-\eta  H(\xi) + \xi  H(\eta)}{\eta-\xi} + \log_2(1+2^{\frac{H(\xi)-H(\eta)}{\eta-\xi}}).
\label{bnas}
\end{equation}
Because this approximation reduces the continuous distribution of outputs to only two choices, ``on'' or ``off'', it can underestimate the true channel capacity and is therefore a lower bound.

\subsection{Exact solution}
The information between the input and output in Eq (\ref{mutinf}) can be maximized numerically for any input/output kernel, $p(g|c)$, if the variables $c$ and $g$ are discretized, making the solution space that needs to be searched, $p(c_i)$, finite. One possibility is to use a gradient descent-based method and make sure that the solution procedure always stays within the domain boundaries $\sum_i p(c_i)=1,p(c_j)\geq 0$ for every $j$. Alternatively, a procedure known as Blahut-Arimoto algorithm has been derived specifically for the purpose of finding optimal channel capacities \cite{blahut}. Both methods yield consistent solutions, but we prefer to use the second one because of faster convergence and convenient inclusion of constraints on the cost of coding (see Appendix \ref{ap:opt} for details).

One should be careful in interpreting the results of such naive optimization and worry about the artifacts introduced by discretization of input and output domains. After discretization, the formal optimal solution is no longer required to be smooth and could, in fact, be composed of a collection of Dirac-delta function spikes. 
On the other hand, 
 the real, physical concentration $c$ cannot be tuned with arbitrary precision in the cell; it is a result of noisy gene expression, and even if this noise source were removed, the \emph{local} concentration at the binding site would still be subject to fluctuations caused by randomness in diffusive flux \cite{bialek+setayeshgar_05,tkacik+al_06}. 
The Blahut-Arimoto algorithm is completely agnostic as to which (physical) concentrations belong to which bins after concentration has been discretized, and so it could assign wildly different probabilities to concentration bins that differ in concentration by less than $\sigma_c$ (i.e. the scale of local concentration fluctuations), making such a naive solution physically unrealizable.
In Appendix \ref{ap:opt} we describe how to properly use Blahut-Arimoto algorithm despite the difficulties induced by discretization. 
\section{A model of signals and noise}
\label{sandn}
If enough data were available, one could directly sample $p(g|c)$ and proceed by calculating the optimal solutions as described previously. Here we start, in contrast, by assuming a Gaussian model of Eq (\ref{gaussianioo}), in which the mean, $\bar{g}(c)$, and the output variance, $\sigma^2_g(c)$, are functions of the transcription factor concentration, $c$. Our goal for this section is to build an effective microscopic model of transcriptional regulation and gene expression, and therefore define both functions with a small number of biologically interpretable parameters. In the subsequent discussion we plan to vary those and thus systematically observe the changes in information capacity.

In the simplest picture, the interaction of the TF with the promoter site consists of binding with a (second order) rate constant $k_+$ and unbinding at a rate $k_-$. In a somewhat more complicated case where $h$ TF molecules cooperatively activate the promoter, the analysis still remains simple as long as the favorable interaction energy between the TFs is sufficient to make only the fully occupied (and thus activated) and the empty (and thus inactivated) states of the promoter likely; this effective two-state system is once more describable with a single rate for switching off the promoter, $k_-$, and the corresponding activation rate has to be  $\propto c^h$  (see Ref \cite{tkacik+al_06}, in particular Appendix B). Generally, therefore, the equilibrium occupancy of the site will be:
\begin{equation}
n=\frac{c^h}{c^h+K_d^h}, \label{meanio}
\end{equation}
where the Hill coefficient, $h$, captures the effects of cooperative binding, and $K_d$ is the equilibrium constant of binding. The mean expression level $g$ is then:
\begin{equation}
g(c)=g_0 \bar{g}=g_0 \,\left\{
\begin{tabular}{ccc}
$n$  & &  $\mathrm{activator}$ \\
$1-n$  & & $\mathrm{repressor}$
\end{tabular}
\right., \label{ftog}
\end{equation}
where $\bar{g}$ has been normalized to vary between 0 and 1, and $g_0$ is the maximum expression level. In what follows we will assume the activator case, where $\bar{g}=n$, and present the result for the repressor at the end.

The fluctuations in occupancy have a (binomial) variance $\sigma_n^2 = n(1-n)$ and a correlation time $\tau_c = 1/(k_+c^h + k_-)$ \cite{tkacik+al_06}.  If the expression level of the target gene is effectively determined by the average of the promoter site occupancy over some window of time $\tau_{\rm int}$, then the contribution to variance in the expression level due to the ``on-off'' promoter switching will be:
\begin{equation}
\left(\frac{\sigma_g}{g_0}\right)^2 =  \sigma_n^2
{{\tau_c}\over{\tau_{\rm int}}}
= \frac{n(1-n)}{(k_+c^h+k_-)\tau_{\rm int}}
  = {{n(1-n)^2}\over{k_- \tau_{\rm int}}} ,
 \label{varf1}
\end{equation}
where in the last step we use the fact that $k_+ c^h(1-n) = k_- n$.  

At low TF concentrations the arrival times of single transcription factor molecules to the binding site are  random events. Recent measurements \cite{gregor+al_06b} seem to be consistent with the hypothesis  that this variability in diffusive flux contributes an additional noise term \cite{bialek+setayeshgar_05,tkacik+al_06}, similar to the Berg-Purcell limit to chemoattractant detection in chemotaxis. The noise in expression level due to fluctuations in the binding site occupancy, or the total {\em input} noise, is therefore a sum of this diffusive component (see Eq (11) of Ref \cite{tkacik+al_06}) and the switching component of Eq (\ref{varf1}):
\begin{equation}
\left(\frac{\sigma_g}{g_0}\right)^2_\mathrm{input}=\frac{n(1-n)^2}{k_-\tau_{\rm int}}+\frac{h^2(1-n)^2 n^2}{\pi D a c\tau_{\rm int}}, \label{inputnoise}
\end{equation}
where $D$ is the diffusion constant for the TF and $a$ is the receptor site size, $a\sim 3\e{nm}$ for a typical binding site on the DNA.

To compute the information capacity in the small noise limit using the simple model developed so far we need the constant $Z$ from Eq (\ref{ZZ}), which is defined as an integral over expression levels.  As both input noise terms are proportional to $(1-\bar{g})^2$, the integral must take the form:
\begin{equation}
Z \propto \int_0^1  {{d \bar{g}}\over{(1-\bar{g})\,F(\bar{g})}},
\end{equation}
where $F(\bar{g})$ is a function that approaches a constant as $\bar{g}\rightarrow 1$.
Strangely, we see that this integral diverges near full induction ($\bar{g}=1$), which means that the information capacity also diverges.

Naively we expect that modulations in transcription factor concentration are {\em not} especially effective at transmitting regulatory information once the relevant binding sites are close to complete occupancy.  More quantitatively, the sensitivity of the site occupancy to changes in TF concentration, $\partial n / \partial c$, vanishes as $n \rightarrow 1$, and hence small changes in TF concentration will have vanishingly small effects.  Our intuition breaks down, however, because in thinking only about the mean occupancy we forget that even very small changes in occupancy could be effective if the noise level is sufficiently small.  As we approach complete saturation, the variance in occupancy decreases, and the correlation time of fluctuations becomes shorter and shorter; together these effects cause the standard deviation as seen through an averaging time $\tau_{\rm int}$ to decrease faster than $\partial n / \partial c$, and this mismatch is the origin of the divergence in information capacity.  Of course the information capacity of a physical system can't really be infinite; there must be an extra source of noise (or reduced sensitivity) that becomes limiting as $n\rightarrow 1$.

The noise in Eq (\ref{inputnoise}) captures only the {\em input} noise, i.e. the noise in the protein level caused by the fluctuations in the occupancy of the binding site. 
In contrast, the {\em output} noise arises even when the occupancy of the binding site is fixed (for example, at full induction), and originates in the stochasticity in transcription and translation. The simplest model postulates that when the activator binding site is occupied with fractional occupancy $n$, mRNA molecules are synthesized in a Poisson process at a rate $R_e$ that generates  $R_e \tau_e n$ mRNA molecules on average during the lifetime of a single mRNA molecule, $\tau_e$. Every message is a template for the production of proteins, which is another Poisson process with rate $R_g$. If the integration time is larger than the lifetime of single mRNA molecules, $\tau_{\rm int}\gg\tau_e$, the mean number of proteins produced is $g = R_g \tau_{\rm int} R_e\tau_e n=g_0 n$, and the variance associated with both Poisson processes is \cite{tkacik+al_06}:
\begin{equation}
\left(\frac{\sigma_g}{g_0}\right)^2_\mathrm{output}=\frac{1+R_g\tau_e}{g_0}\,n, \label{noise-output}
\end{equation}
where $b=R_g\tau_e$ is the burst size, or the number of proteins synthesized per mRNA.

We can finally put the results together by adding the input noise Eq (\ref{inputnoise}) and the output noise Eq (\ref{noise-output}), and expressing both in terms of the normalized expression level $\bar{g}(c)$: 
\begin{eqnarray}
\left(\frac{\sigma_g}{g_0}\right)^2_\mathrm{act}&=&\alpha\,\bar{g}+ \nonumber\\
&+&
\beta (1-\bar{g})^{2+\frac{1}{h}} \bar{g}^{2-\frac{1}{h}} +\gamma\bar{g}(1-\bar{g})^2,\label{actnoise}\\
\left(\frac{\sigma_g}{g_0}\right)^2_\mathrm{rep}&=&\alpha\,\bar{g}+\nonumber\\
&+&
\beta (1-\bar{g})^{2-\frac{1}{h}} \bar{g}^{2+\frac{1}{h}} + \gamma\bar{g}^2(1-\bar{g}), \label{repnoise}
\end{eqnarray}
with the relevant parameters $\left\{\alpha,\beta,\gamma,h\right\}$ explained in Table \ref{t-params}. Note that both repressor and activator cases differ only in the shape of the input noise contributions (especially for low cooperativity $h$). Note further that the output noise increases monotonically with mean expression $\bar{g}$, while the input noise peaks at the intermediate levels of expression. To make the examination of the parameter space in the next section feasible, we set $\gamma=0$; models with switching noise instead of diffusive noise produce qualitatively similar results. 
\begin{table}
\centering
\begin{tabular}{|l|c|l|}
\hline
Parameter	&	Value	&	 Description	\\ \hline\hline
$\alpha$		& $(1+b)/g_0$	&	Output noise strength	\\ \hline
$\beta$		& $h^2/\pi D a K_d \tau_{\rm int}$ &Diffusion input noise strength	\\\hline
$\gamma$	& $(k_-\tau_{\rm int})^{-1}$&	Switching input noise strength \\ \hline
$h$			& 		& Cooperativity (Hill coefficient) \\ \hline
\end{tabular}
\caption[Noise model parameters]{Gaussian noise model parameters. Note that if burst size $b\gg 1$, then the output noise is determined by the average number of mRNA molecules, $\alpha \sim (\langle\mathrm{mRNA}\rangle)^{-1}$. Note further that if the on-rate is diffusion limited, i.e. $k_+=4\pi D a$, then both input noise magnitudes, $\beta$ and $\gamma$, are proportional and decrease with increasing $k_-$, or alternatively, with increasing $K_d=k_-/k_+$. }
\label{t-params}
\end{table}
\section{Results}
\subsection{Capacity of simple regulatory elements}
\begin{figure} 
   \centering
   \includegraphics[width=3.3in]{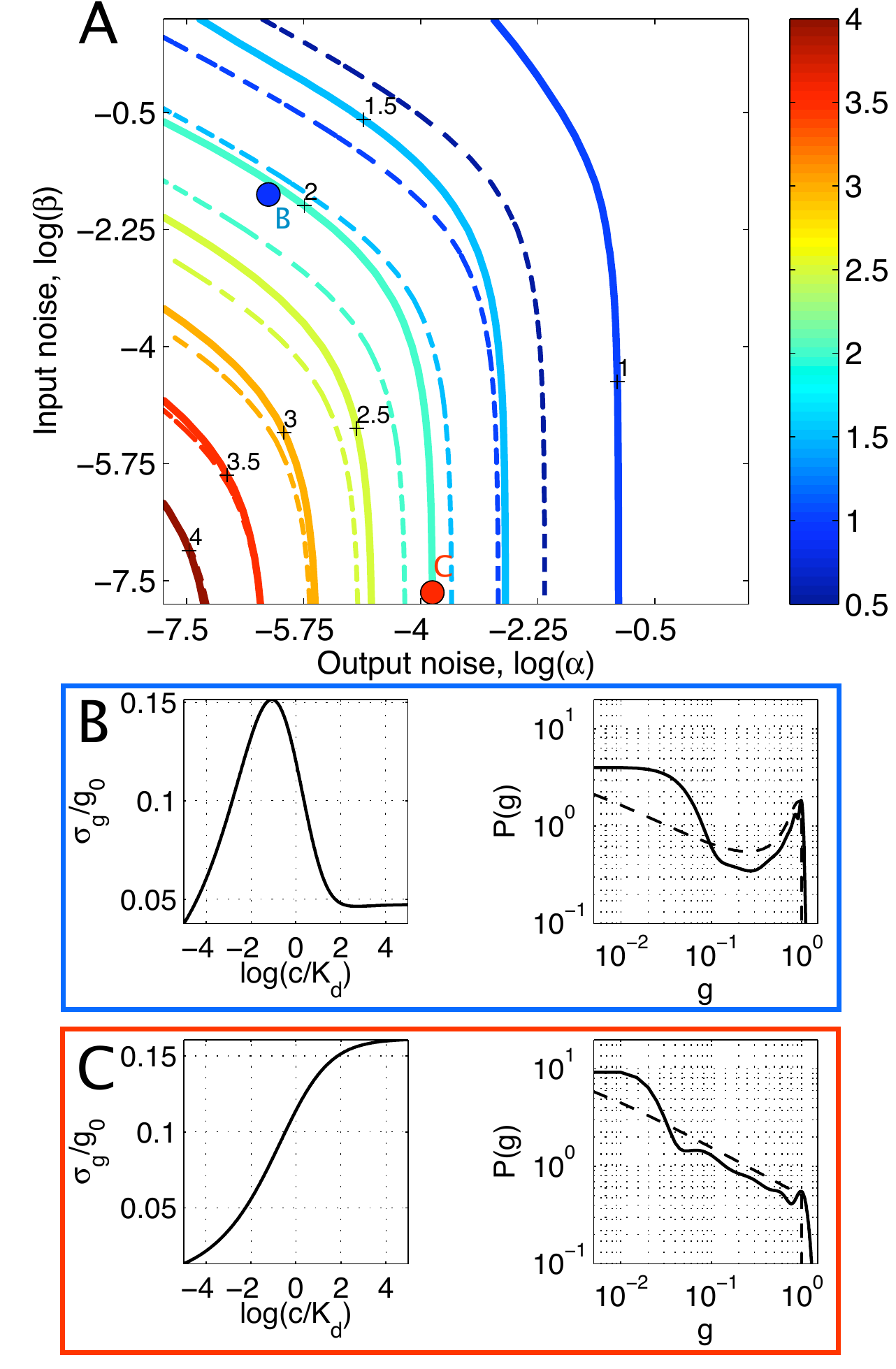} 
   \caption{(Color online) Information capacity (color code, in bits) as a function of input and output noise using the activator input/output relation with Gaussian noise given by Eq (\ref{exnoise}) and no cooperativity ($h=1$). Panel A shows the exact capacity calculation (thick line) and the small noise approximation (dashed line). Panel B displays the details of the blue point in A: the noise in the output is shown as a function of the input, with a peak being characteristic of a dominant input noise contribution; also shown is the exact solution (thick black line) and the small-noise approximation (dashed black line) to the optimal distribution of output expression levels. Panel C similarly displays details of the system denoted by a red dot in A; here the output noise is dominant and both approximate and exact solutions for the optimal distribution of outputs show a trend monotonically decreasing with the mean output.  } 
   \label{f-cpanel1}
\end{figure}

Having at our disposal both a simple model of signals and noise and a numerical way of finding the optimal solutions given an arbitrary input/output kernel, we are now ready to examine the channel capacity as a function of the noise parameters from Table \ref{t-params}.
Our first result, shown in Fig \ref{f-cpanel1}, concerns the simplest case of an activator with no cooperativity, $h=1$; for this case, the noise in Eq (\ref{actnoise}) simplifies to:
\begin{equation}
\left(\frac{\sigma_g}{{g_0}}\right)^2=\alpha \bar{g} + \beta (1-\bar{g})^3 \bar{g}. \label{exnoise}
\end{equation}
Here we have assumed that there are two relevant sources of noise, i.e. the output noise (which we parametrize by $\alpha$ and plot on the horizontal axis) and the input diffusion noise (parametrized by $\beta$, vertical axis). Each point of the noise plane in Fig \ref{f-cpanel1}a therefore represents a system characterized by a Gaussian noise model, Eq (\ref{gaussianioo}), with variance given by Eq (\ref{exnoise}) above.

As expected, the capacity increases most rapidly when the origin of the noise plane is approached approximately along its diagonal, whereas along each of the edges one of the two noise sources effectively disappears, leaving the system dominated by either output or input noise alone. 
We pick two illustrative examples, the blue and the red systems of  Figs \ref{f-cpanel1}b and \ref{f-cpanel1}c, that have realistic noise parameters. The blue system has, apart for the decreased cooperativity ($h=1$ instead of $h=5$), the characteristics of the Bicoid-Hunchback regulatory element in \emph{Drosophila melanogaster} \cite{tkacik+al_06,tkacik+al_07}; the red system is dominated by output noise with characteristics measured recently for about 40 yeast genes \cite{bar-even+al_06}. We would like to emphasize that  both the small-noise approximation and the exact solution predict that these realistic systems are capable of transmitting more than 1 bit of regulatory information and that they, indeed, could transmit up to about 2 bits. In addition, we are also reminded that while the distributions (for example the optimal output distribution in Fig \ref{f-cpanel1}b) can look bimodal and this has often been taken as an indication that there are two relevant states of the output, such distributions really can have capacities above 1 bit; similarly, distributions without prominent features, such as monotonically decreasing optimal output distribution of Fig \ref{f-cpanel1}c, should also not be  expected necessarily to have low capacities.

A closer look at the overall agreement between the small-noise approximation (dashed lines in Fig \ref{f-cpanel1}a) and the exact solution (thick lines) shows that the small-noise approximation underestimates the true capacity, consistent with our remark that for large noise the approximation will incorrectly produce negative results; at the 2-bit information contour the approximation is about $\sim 15\%$ off but improves as the capacity is increased. 

In the high noise regime we are making yet another approximation, the validity of which we now need to examine. In our discussion about the models of signals and noise we assumed that we can talk about the fractional occupancy of the binding site and the \emph{continuous} concentrations of mRNA, transcription factors and protein, instead of counting these species in discrete units, and that noise can effectively be treated as Gaussian. Both of these assumptions are the cornerstones of the Langevin approximation for calculating the noise variance \cite{vanKampen_07}. If parameters $\alpha$ and $\beta$  actually arise due to the underlying microscopic mechanisms described in the Section \ref{sandn} on signals and noise, we expect that at least for some large-noise regions of the noise plane the discreteness in the number of mRNA molecules will become important and the Langevin approximation will fail. In such cases (a much more time-consuming) exact calculation of the input/output relations using the Master equation is possible for some noise models (see Appendix \ref{ax:lan});  we show that in the region where $\log\alpha>-2$ the channel capacities calculated with Gaussian kernels can be overestimated by $\sim 10\%$ or more; there the Langevin calculation gives the correct second moment, but misses the true shape of the distribution. Although both examples with realistic noise parameters, B and C, of Fig \ref{f-cpanel1} lie safely in the region where Langevin approximation is valid, care should be used whenever both output noise $\alpha$ and burst size are large, and $\alpha$ is consequently dominated by small number of transcripts.

\begin{figure} 
   \centering
   \includegraphics[width=3.5in]{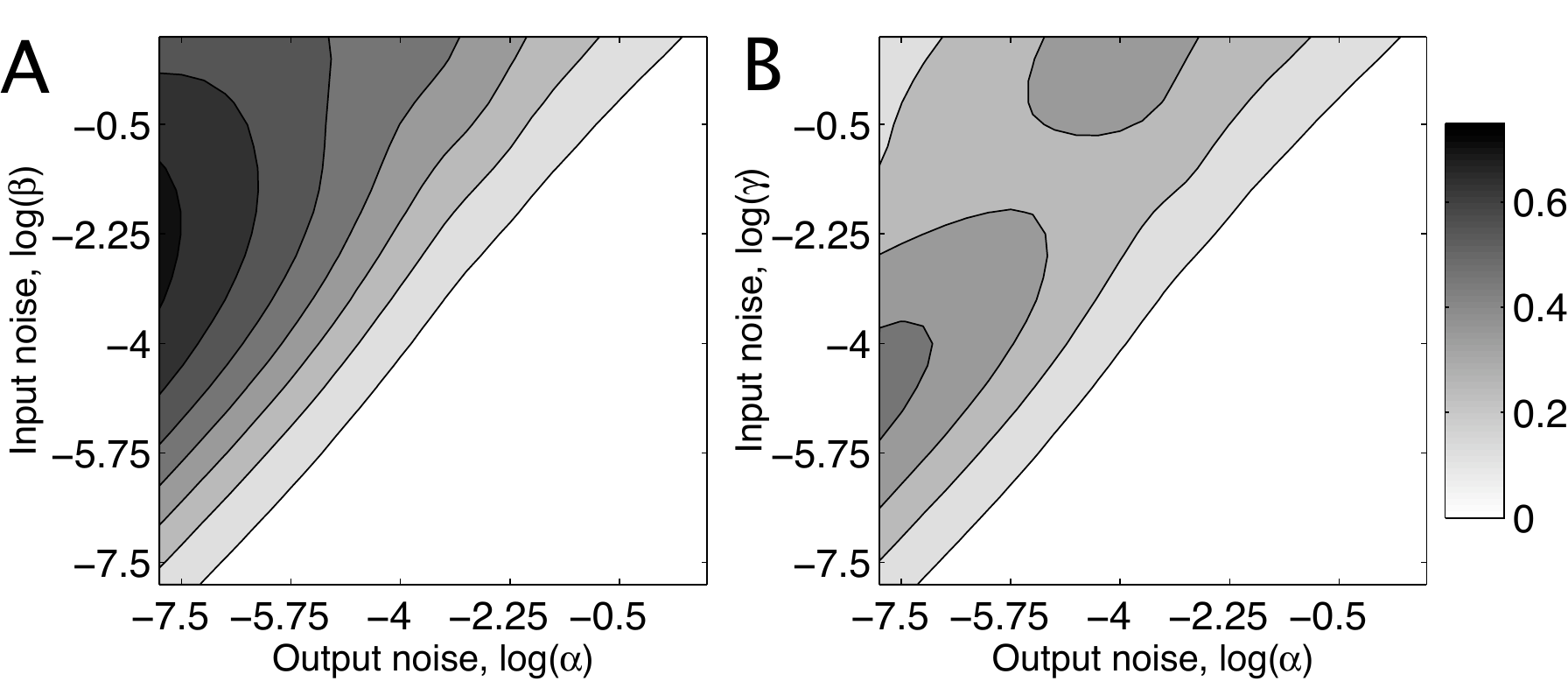} 
   \caption{Difference in the information capacity between the repressors and activators (color code in bits). Panel A shows $I_{\rm rep}(h=1)-I_{\rm act}(h=1)$, with the noise model that includes output ($\alpha$) and input diffusion noise ($\beta$) contributions (see Fig \ref{f-cpanel1} for absolute values of $I_{\rm act} (h=1)$). Panel B shows $I_{\rm rep}-I_{\rm act}$ for the noise model  that includes output noise ($\alpha$) and input switching noise ($\gamma$) contributions; this plot is independent of cooperativity, $h$.}
   \label{f-actrep}
   \end{figure}

\begin{figure} 
   \centering
   \includegraphics[width=3.5in]{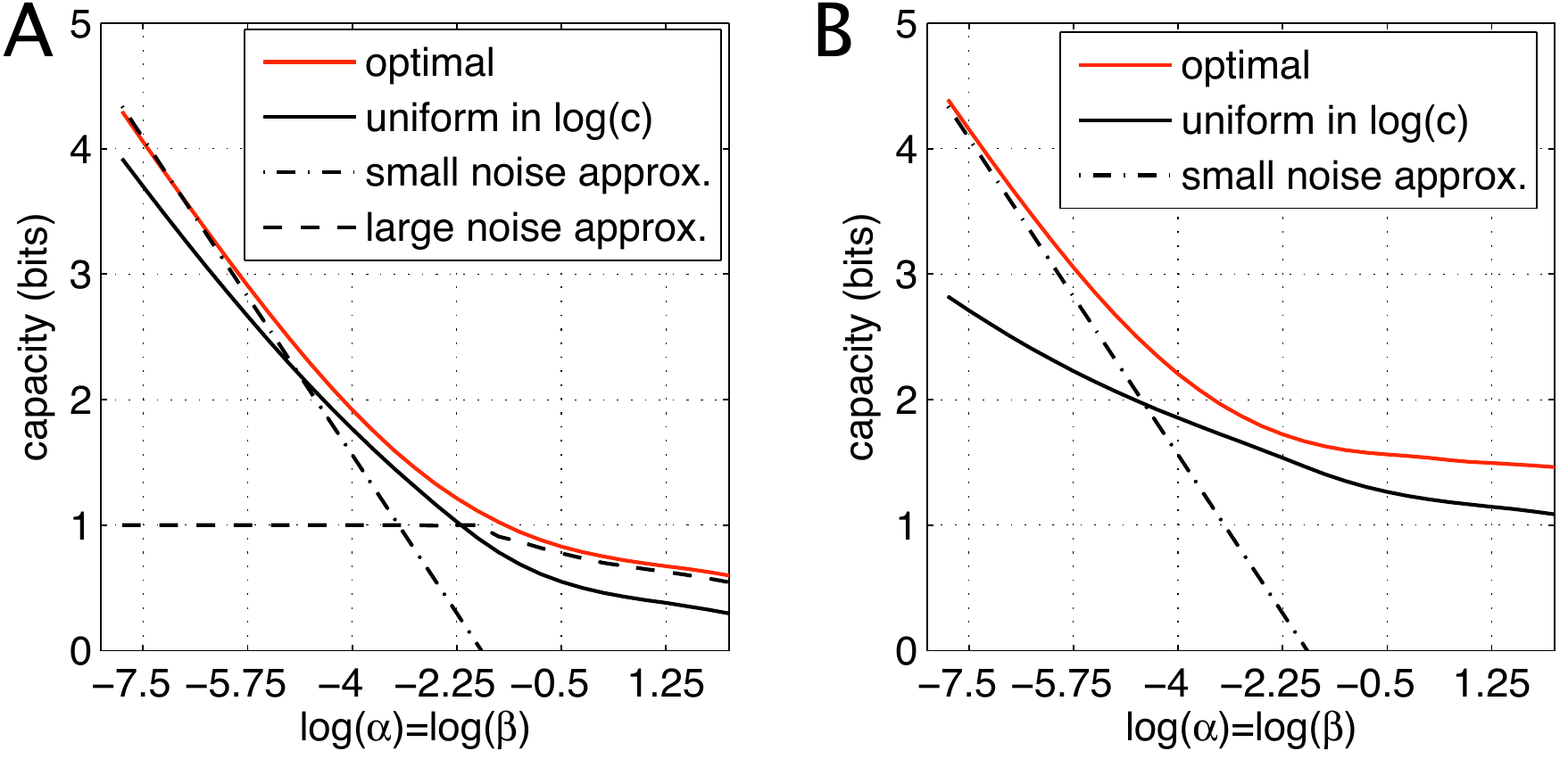} 
   \caption{(Color online) Comparison of exact channel capacities and various approximate solutions. For both panels (panel A, no cooperativity, $h=1$; panel B, strong cooperativity, $h=3$) we take a cross-section through the noise plane in Fig \ref{f-cpanel1} along the main diagonal, where the values for noise strength parameters $\alpha$ and $\beta$ are equal. The exact optimal solution is shown in red. By moving along the diagonal of the noise plane (and along the horizontal axis in the plots above) one changes both input and output noise by the same multiplicative factor $s$, and since, in small-noise approximation, $I_{\rm SNA}\propto \log Z$, $Z =\int \sigma_g(\bar{g})^{-1}\,d\bar{g}$, that factor results in an additive change in capacity by $\log_2 s$. We can use the large noise approximation lower bound on capacity for the case $h=1$, in the parameter region where capacities fall below 1 bit.} 
   \label{f-axs}
\end{figure}

Is there any difference between activators and repressors in their capacity to convey information about the input? We concluded Section \ref{sandn} on the noise models with separate expressions for activator noise, Eq (\ref{actnoise}), and repressor noise, Eq (\ref{repnoise}); focusing now on the repressor case, we recompute the information in the same manner as we did for the activator in Fig \ref{f-cpanel1}a, and display the difference between the capacities of the repressor and activator with the same noise parameters in Fig \ref{f-actrep}. As expected, the biggest difference occurs above the main diagonal, where the input noise dominates over the output noise. In this region the capacity of the repressor can be bigger by as much as third  than that of the corresponding activator. Note that as $h\rightarrow\infty$, the activator and repressor noise expressions become indistinguishable and the difference in capacity vanishes for the noise models with output and input diffusion noise contributions, Eqs (\ref{actnoise}, \ref{repnoise}). 
The behavior of the regulatory element is conveniently visualized in Fig \ref{f-axs} by plotting a cut through the noise plane along its main diagonal. Moving along this cut scales the total noise variance of the system up or down by a multiplicative factor, and allows us to observe the overall agreement between the exact solution and small- and large-noise approximations.
In addition we point out the following interesting features of Fig \ref{f-axs} that will be examined more closely in subsequent sections.

First, the parameter region in non-cooperative case, in which the capacity falls below one bit and the large noise approximation is applicable, is small and shrinks further at higher cooperativity. This suggests that a biological implementation of a reliable binary channel could be relatively straightforward, assuming our noise models are appropriate. Moreover, there exist distributions not specifically optimized for the input/output kernel, such as the input distribution uniform in $\log (c/K_d)$  that we pick as illustrative example in Fig \ref{f-axs} (thick black line); we find that this simple choice can achieve considerable information transmission, and are therefore motivated to raise a more general question about the sensitivity of channel capacity with respect to perturbations in the optimal solution, $p^*(c)$. We revisit this idea more systematically in the next section. 

Second, it can be seen from Fig \ref{f-axs} that at small noise the cooperativity has a minor effect on the channel capacity.  This is perhaps unexpected as the shape of the mean response $\bar{g}(c)$ strongly depends on $h$. We recall, however, that mutual information $I(c;g)$ is invariant to any invertible reparametrization of either $g$ or $c$. In particular, changing the cooperativity or the  value of the equilibrium binding constant, $K_d$, in theory only results in an invertible change in the input variable $c$, and therefore the change in the steepness or midpoint of the mean response must not have any effect on $I(c;g)$. 
This argument does break down in the high noise regime, where the cooperative system achieves capacities above one bit while the non-cooperative system fails to do so. Reparametrization invariance would work only if the input concentration could extend over the whole positive interval, from zero to infinity. The substantial difference between capacities of cooperative and non-cooperative systems in Fig \ref{f-axs} at \emph{low capacity} stems from the fact that in reality the cell (and our computation) is limited to a finite range of concentrations, $c\in\left[c_{\rm min},c_{\rm max}\right]$, instead of the whole positive half-axis, $c\in\left[0,\infty\right)$ . We explore the issue of limited input dynamic range further in the next section. 

Finally, we draw attention to the simple linear scaling of the channel capacity with the logarithm of the total noise strength in small noise approximation, as explained in the caption of Fig \ref{f-axs}. In general, increasing the number of input and output molecules by a factor of four will decrease the relative input and output noise by a factor of $\sqrt{4}=2$, and therefore, in the small noise approximation, increase the capacity by $\log_2 2=1 \e{bit}$. If one assumes that the cell can make transcription factor and output protein molecules at no cost, then scaling of the noise variance along the horizontal axis of Fig \ref{f-axs} is inversely proportional to the total number of signaling molecules used by the regulatory element, and its capacity can grow without bound as more and more signaling molecules are used. If, however, there are metabolic or time costs to making more molecules, our optimization needs to be modified appropriately, and we present the relevant computation in Section \ref{scosts} on the costs of coding.
\subsection{Cooperativity, dynamic range and the tuning of solutions}

In the analysis presented so far we have not paid any particular attention to the question of whether the optimal input distributions are biologically realizable or not. We will proceed to relax some of the idealizations made until now and analyze the corresponding changes in the information capacity.

\begin{figure} 
   \centering
   \includegraphics[width=3.5in]{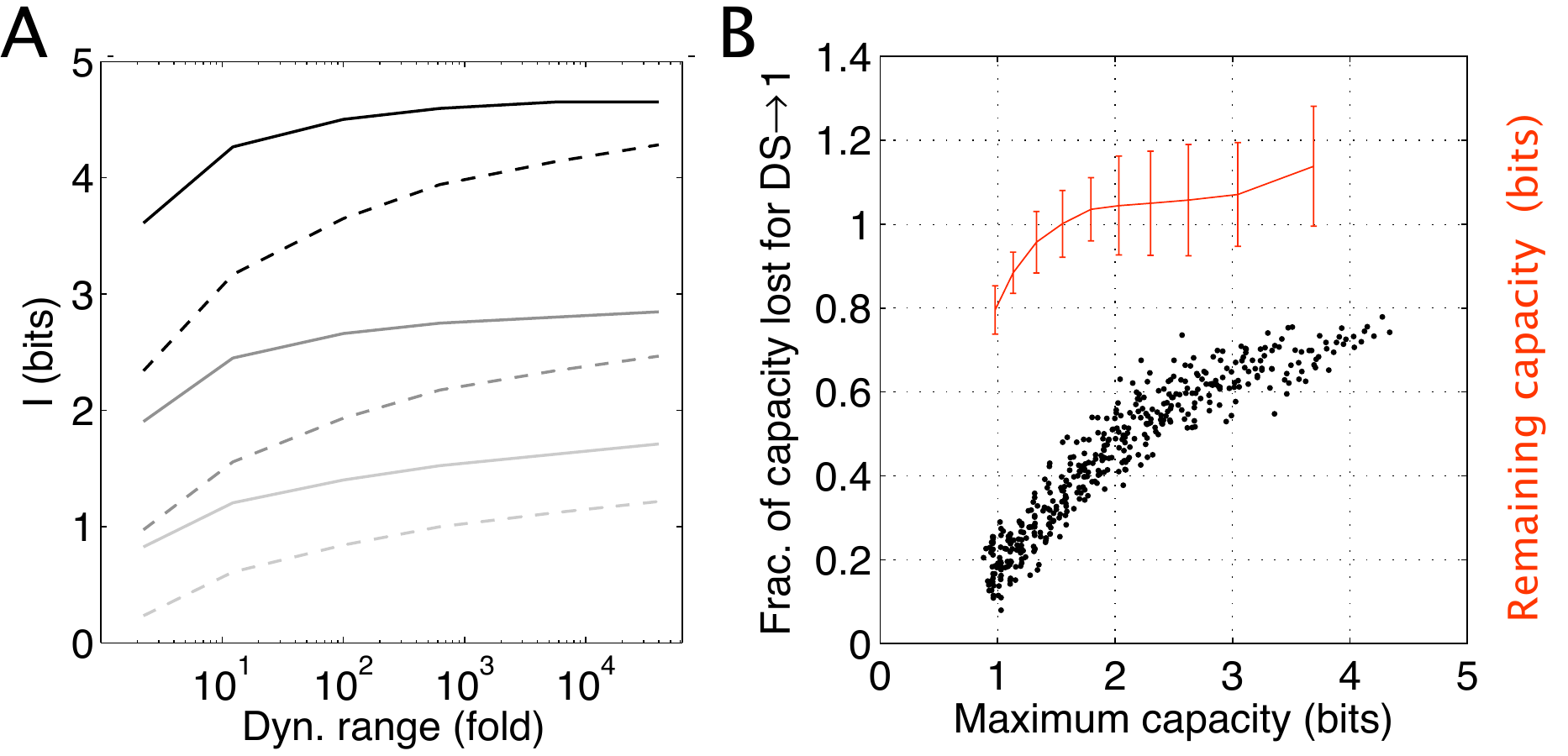} 
   \caption{(Color online) Effects of imposing realistic constraints on the space of allowed input distributions. Panel A shows the change in capacity if the dynamic range of the input around $K_d$ is changed (``25-fold range'' means $c\in\left[\frac{1}{5}K_d,5K_d\right]$). The regulatory element is a repressor with either no cooperativity (dashed line) or high cooperativity, $h=3$ (thick line). We plot three high-low cooperativity pairs for different choices of the output noise magnitude (high noise in light gray, $\log\alpha\approx -2.5$; medium noise in dark gray, $\log\alpha\approx -5$; low noise in black, $\log\alpha\approx -7.5$). Panel B shows the sensitivity of channel capacity to perturbations in the optimal input distribution. For various systems from Fig \ref{f-cpanel1} we construct suboptimal input distributions, as described in the text, compute the fraction of capacity lost relative to the unperturbed optimal solution and plot this fraction against the optimal capacity of that system (black dots); extrapolated absolute capacity left when the input tends to be very different from optimal, i.e. $D_{\rm JS}\rightarrow 1$, is plotted in red.  } 
   \label{f-costs_x}
\end{figure}

We start by considering the impact on channel capacity of changing the allowed dynamic range to which the input concentration is restricted. Figure \ref{f-costs_x}a displays the capacity as a function of the dynamic range, output noise and cooperativity. The main feature of the plot is the difference between the low and high cooperativity cases at each noise level; regardless of cooperativity the total information at infinite dynamic range would saturate at approximately the same value (which depends on the output noise magnitude). However, highly cooperative systems manage to reach a high fraction of $80\%$ or more of their saturated information capacity even at reasonable dynamic ranges of 25 to 100-fold (meaning that the input concentration varies between $[\frac{1}{5}K_d,5K_d]$ or $[\frac{1}{10}K_d,10K_d]$, respectively), whereas low cooperativity systems require a much bigger dynamic range for the same effect. The decrease in capacity with decreasing dynamic range is a direct consequence of the nonlinear relationship between the concentration and occupancy, Eq (\ref{meanio}), and for low cooperativity systems means being unable to fully shut down or fully induce the promoter. In theory, Eq (\ref{actnoise}) predicts that $\sigma_g(\bar{g}(c)\rightarrow 0)= 0$, making the state in which the gene is ``off'' very informative about the input. If, however, the gene cannot be fully repressed {either} because there is always some residual input, $c_{\rm min}$, {or} because there is leaky expression even when the input is exactly zero, then at any biologically reasonable input dynamic range some capacity will be lost.

Next, we briefly discuss how precisely tuned the resulting optimal distributions have to be to take full advantage of the regulatory element's capacity. For each point in the noise plane of Fig \ref{f-cpanel1}a the optimal input distribution $p^*(c)$ is perturbed many times to create an ensemble of suboptimal inputs $p_i(c)$ (see Appendix \ref{ax:pert}). For each $p_i(c)$, we compute, first, its distance away from the optimal solution by means of Jensen-Shannon divergence, $d_i=D_{\rm JS}(p_i,p^*)$ \cite{lin_91}; next, we use the $p_i(c)$ to compute the suboptimal information transmission $I_i$. 
The divergence $d_i$ is a measure of similarity between two distributions and ranges between 0 (distributions are the same) and 1 (distributions are very different); $1/d_i(p_i,p^*)$ approximately corresponds to the number of samples one would have to draw to say with confidence that they were selected either from $p_i$ or $p^*$.
A scatter plot of many such pairs $(d_i,I_i)$ obtained with various perturbations $p_i(c)$ for each system of the noise plane characterizes the sensitivity of the optimal solution for that system; the main feature of such a plot, Fig \ref{f-optperturb}, is the linear (negative) slope that describes the fraction of channel capacity lost for a unit of Jensen-Shannon distance away from the optimal solution. Figure \ref{f-costs_x}b displays these fractions as a function of the optimal capacity, and each system from the noise plane shown in Fig \ref{f-cpanel1} is represented by a black dot. We note that systems with higher capacities require more finely tuned solutions and suffer a larger fractional (and thus absolute) loss if the optimal input distribution is perturbed.  Importantly, if the linear slopes are taken seriously and are used to extrapolate towards distributions that are very different from optimal, $D_{\rm JS}\rightarrow 1$,  we observe that for most of the noise plane the leftover capacity still remains about a bit, indicating that biological regulatory elements capable of transmitting an ``on-off'' decision perhaps are not difficult to construct. On the other hand, transmitting significantly more than one bit requires some degree of tuning that matches the distribution of inputs to the characteristics of the regulatory element.

\subsection{Costs of higher capacity}
\label{scosts}
Real regulatory elements must balance the pressure to convey information reliably with the cost of maintaining the cell's internal state, represented by the expression levels of transcription factors. The fidelity of the representation is increased  (and the fractional fluctuation in their number is decreased) by having more molecules ``encode''  a given state. On the other hand, making or degrading more transcription factors puts a metabolic burden on the cell, and frequent transitions between various regulatory states could involve large time lags as, for example, the regulation machinery attempts to keep up with a changed environmental condition, by accumulating or degrading the corresponding TF molecules. In addition, the output genes themselves that get switched on or off by transcription factors and therefore ``read out'' the internal state must not be too noisy, otherwise the advantage of maintaining precise transcription factor levels is lost.

Suppose that there is a cost to the cell for each molecule of output gene that it needs to produce, and that this incremental cost per molecule is independent of the number of molecules already present. 
Then, on the output side, the cost must be proportional to $\langle g \rangle=\int dg\, g \, p(g)$. We remember that in optimal distribution calculations $g$ is expressed as relative to the maximal expression, such that its mean is between zero and one. To get an absolute cost in terms of the number of molecules, this normalized $\bar{g}$ therefore needs to be multiplied by the inverse of the output noise strength, $\alpha^{-1}$, as the latter scales with $g_0$ (see Table \ref{t-params}). The contribution of the output cost is thus $\propto \alpha^{-1} \bar{g}$.

On the input side, the situation is similar: the cost must be proportional to $K_d \langle \tilde{c}\rangle=K_d \int d{\tilde{c}}\, \tilde{c} \,p(\tilde{c})$, where our optimal solutions are expressed, as usual, in dimensionless concentration units, $\tilde{c}=c/K_d$. In either of the two input noise models (i.e. diffusion or switching input noise), with diffusion constant held fixed, $K_d\propto \beta^{-1}$ or $K_d\propto \gamma^{-1}$. See Appendix \ref{ax:nb} for the notes on the effects of non-specific binding of transcription factors to the DNA.

Collecting all our thoughts on the costs of coding, we can write down the ``cost functional'' as the sum of input and output cost contributions:
\begin{equation}
\langle \mathcal{C}[p(c)]\rangle=\frac{v_1}{\beta} \int dc\, p(c)\, c + \frac{v_2}{\alpha} \int dc\, p(c)\, \int dg\, p(g|c)\, g, \label{cconstraint}
\end{equation}
where $v_1$ and $v_2$ are proportional to the unknown costs per molecule of input or output, respectively, and $\alpha$ and $\beta$ are noise parameters of Table \ref{t-params}. This ansatz captures the intuition that while decreasing noise strengths will increase information transmission, it will also increase the cost. Instead of maximizing the information without regard to the cost, the new problem to extremize is:
\begin{eqnarray}
\mathcal{L}[p(c)]&=& I[p(c)] - \Phi \langle\mathcal{C}[p(c)]\rangle - \Lambda \int dc\,p(c), 
\end{eqnarray}
and the Lagrange multiplier $\Phi$ has to be chosen so that the cost of the resulting optimal solution $\langle \mathcal{C}[p^*(c)]\rangle$ equals some predefined cost $C_0$ that the cell is prepared to pay. 
\begin{figure} 
   \centering
   \includegraphics[width=3.5in]{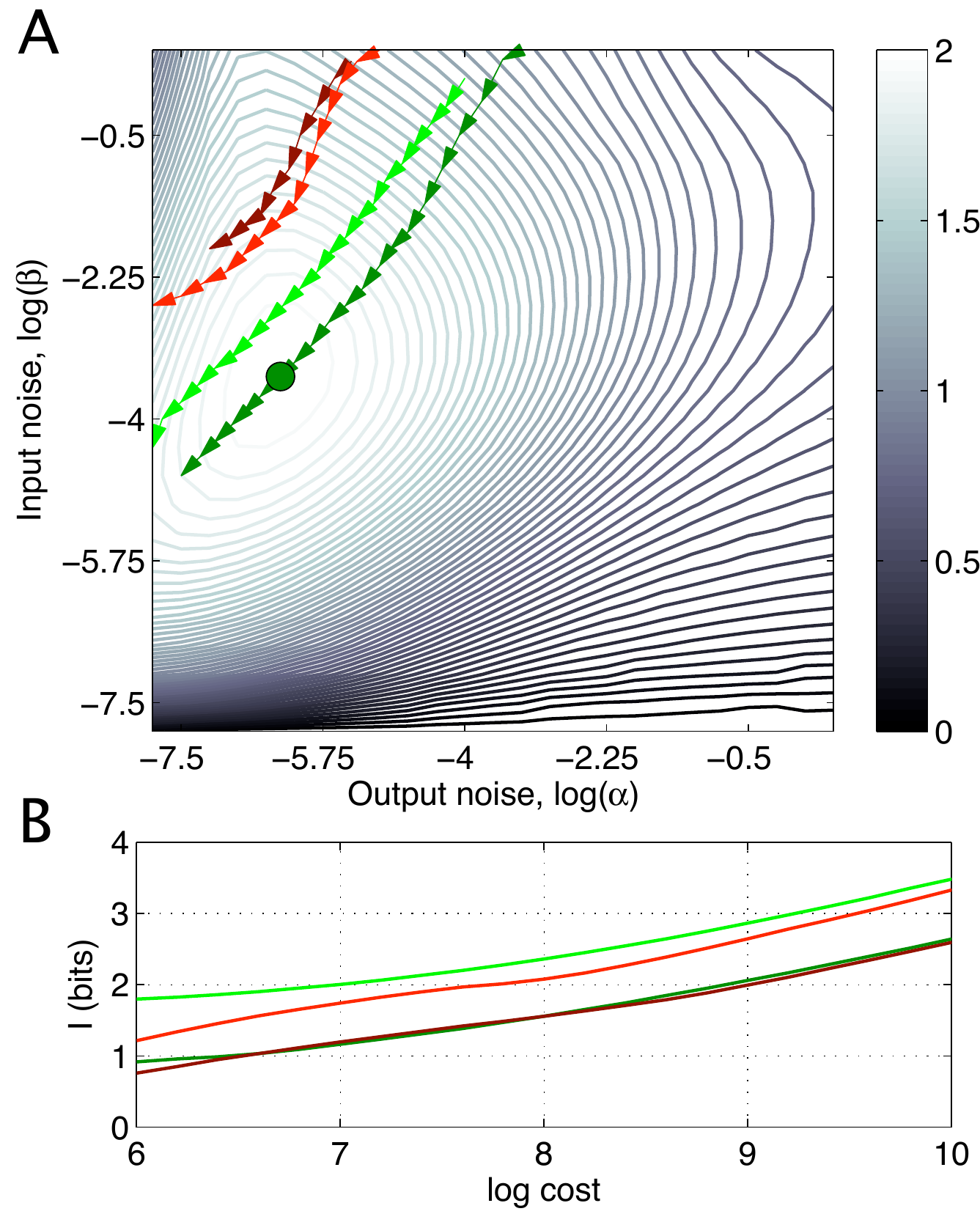} 
   \caption{(Color online) The effects of metabolic or time costs on the achievable capacity of simple regulatory elements. Contours in panel A show the noise plane for non-cooperative activator from Fig \ref{f-cpanel1}, with the imposed constraint that the average total (input + output) cost is fixed to some $C_0$; as the cost is increased, the optimal solution (green dot) moves along the arrows on a dark green line (the contours change correspondingly, not shown). Light green line shows activator with cooperativity $h=3$, dark and light red lines show repressors without and with cooperativity ($h=3$). Panel B shows the achievable capacity as a function of cost for each line in panel A.} 
   \label{f-costs}
\end{figure}

We now wish to recreate the noise plane of Fig \ref{f-cpanel1}, while constraining the total cost of each solution to $C_0$. To be concrete and pick the value for the cost and proportionality constants in Eq (\ref{cconstraint}), we use the estimates from \emph{Drosophila} noise measurements and analysis in \cite{gregor+al_06b, tkacik+al_06}, which assign to the system denoted by a blue dot in Fig \ref{f-cpanel1}a, the values of $\sim 800$ Bicoid molecules of input at $K_d$, and a maximal induction of $g_0\sim 4000$ Hunchback molecules if the burst size $b$ were 10. Figure \ref{f-costs}a is the noise plane for an activator with no cooperativity, as in Fig \ref{f-cpanel1}, but with the cost limited to an average total of $C_0 \sim 7000$ molecules of input and output per nucleus. There is now one optimal solution denoted by a green dot (with a dominant input noise contribution); if one tries to choose a system with lower input or output noise, the cost constraint forces the input distribution, $p(c)$, and the output distribution, $p(g)$, to have very low probabilities at high induction, consequently limiting the capacity. 

Clearly, a different system will be optimal if another total allowed cost $C_0$ is selected. The dark green line on the noise plane in Fig \ref{f-costs}a corresponds to the flow of the optimal solution for an activator with no cooperativity if the allowed cost is increased, and the corresponding cost-capacity curve is shown in Fig \ref{f-costs}b. The light green line is the trajectory of the optimal solution in the noise plane of the activator system with cooperativity $h=3$, and the dark and light red trajectories are shown for the repressor with $h=1$ and $h=3$, respectively. We note first that the behavior of the cost function is quite different for the activator (where low input implies low output and therefore low cost; and conversely high input means high output and also high cost) and the repressor (where input and output are mutually exclusively high or low and the cost is intermediate in both cases). Secondly, in Fig \ref{f-costs}b we observe that the optimal capacity as a function of cost is similar for the activators and repressors, in contrast to the comparison of Fig \ref{f-actrep}, where repressors provided higher capacities. Thirdly, we note in the same figure that increasing the cooperativity at fixed noise strength $\beta$ brings a substantial increase, of almost a bit over the whole cost range, in the channel capacity, in agreement with our previous observations about the interaction between capacity and the dynamic range. The last and perhaps the most significant conclusion is that even with input distributions matched to maximize the transmission at a fixed cost, the capacity still only scales roughly linearly with the logarithm of the number of available signaling molecules, and this fact must ultimately be limiting in a single regulatory element.
\section{Discussion}
We have tried to analyze a simple regulatory element as an information processing device. One of our major results is that one cannot discuss an element in isolation from the statistics of the input that it is exposed to. Yet in cells the inputs are often transcription factor concentrations that ``encode'' the state of various genetic switches, from those responsible for cellular identity to those that control the rates of metabolism and cell division, and the cell exerts control over these concentrations. While it \emph{could} use different distributions to represent various regulatory settings, we argue that the cell \emph{should} use the one distribution that allows it to make the most of its genetic circuitry -- the distribution that maximizes the dependency, or mutual information, between inputs and outputs. Mutual information can then be seen both as a measure of how well the cell is doing by using \emph{its} encoding scheme, and the best it could have done using the \emph{optimal} scheme, which we can compute; comparison between the optimal and measured distributions gives us a sense of how close the organism is to the achievable bound \cite{tkacik+al_07}. Moreover, mutual information has absolute units, i.e. bits, that have a clear interpretation in terms of the number of discrete distinguishable states that the regulatory element can resolve. This last fact helps clarify the ongoing debates about what is the proper noise measure for genetic circuits, and in what context a certain noise is either ``big'' or ``small'' (as it is really a function of the inputs). Information does not replace the standard \emph{noise-over-the-mean} measure -- noise calculations or measurements are still necessary to compute the element's capacity -- but does give it a functional interpretation.

We have considered a class of simple parametrizations of signals and noise that can be used to fit  measurements for several model systems, such as Bicoid-Hunchback in the fruit fly, a number of yeast genes, and the \emph{lac} repressor in \emph{Escherichia coli} (see Ref \cite{tkacik+al_07} for the latter). We find that the capacities of these realistic elements are generally larger than 1 bit, and can be as high as 2 bits. By simple inspection of optimal output distributions in Figs \ref{f-cpanel1}b or \ref{f-cpanel1}c it is difficult to say anything about the capacity: the distribution might look bimodal yet carry more than one bit, or might even be a monotonic function without any obvious structure, indicating that the information is encoded in the graded response of the element. When the noise is sufficiently high, on the other hand, the optimal strategy is that of achieving one bit of capacity and only utilizing maximum and minimum available levels of transcription factors for signaling. The set of distributions that achieve capacities close to the optimal one is large, suggesting that perhaps one-bit switches are not difficult to implement biologically, while in contrast we find that transmission of much more than one bit requires some tuning of the system.

Finally, we discussed how additional biophysical constraints can modify the channel capacity. By assuming a linear cost model for signaling molecules and a limited input dynamic range, the capacity and cost couple in an interesting way and the maximization principle allows new questions to be asked. For example, increasing the cooperativity reduces the cost, as we have shown; on the other hand, it increases the sensitivity to fluctuations in the input, because the input noise strength $\beta$ is proportional to the square of Hill's coefficient, $h^2$. In a given system we could therefore predict the optimal effective cooperativity, if we knew the real cost per molecule. Further work is needed to tease out the consequences of cost (if any) from experimental data.

The principle of information maximization clearly is not the only possible lens through which regulatory networks are to be viewed. One can think of examples where  there are constraints on the \emph{dynamics}, something that our analysis has ignored by only looking at steady state behavior; for instance, the chemotactic network of \emph{Escherichia coli} has to perfectly adapt in order for the bacterium to be able to climb the attractant gradients. Alternatively, suppose that a system needs to convey only a single bit, but it has to be done reliably in a fluctuating environment, perhaps by being \emph{robust} to the changes in outside temperature. In this case it seems that both concepts, that of maximal information transmission and the robustness to fluctuations in certain auxiliary variables which also influence the noise, could be included into the same framework, but the issue needs further work. More generally, however, these and similar examples assume that one has identified in advance the biologically relevant features of the system, e.g. \emph{perfect adaptation} or \emph{robustness}, and that there exists a problem-specific error measure which the regulatory network is trying to minimize. Such a measure could then either replace or complement  the assumption-free information theoretic approach presented here.

We emphasize that the kind of analysis carried out here is not restricted to a single regulatory element. As was pointed out in the introduction, the inputs $\mathcal{I}$ and the outputs $\mathcal{O}$ of the regulatory module can be multi-dimensional, and the module could implement complex internal logic with multiple feedback loops. It seems that especially in such cases, when our intuition about the noise -- now a function of multiple variables -- starts breaking down, the information formalism could prove to be helpful. Although the solution space that needs to be searched in the optimization problem grows exponentially in the inputs, there are biologically relevant situations that nevertheless appear tractable: for example, when there are multiple readouts of the same input, or combinatorial regulation of a single output by a pair of inputs; in addition, knowing that the capacities of a single input/output chain are on the order of a few bits also means that only a small number of distinct input levels for each input need to be considered. Some cases of interest therefore appear immediately amenable to biophysical modeling approaches and the computation of channel capacities, as presented in this paper.

We have focused here on the theoretical exploration of information capacity in simple models. It is natural to ask how our results relate to experiment. Perhaps the strongest connection would be if biological systems really were selected by evolution to optimize information flow in the sense we have discussed. If this optimization principle applies to real regulatory elements, then, for example, given measurements on the input/output relation and noise in the system we can make parameter free predictions for the distribution of expression levels that cells will use. Initial efforts in this direction, using the Bicoid-Hunchback element in the {\em Drosophila} embryo as an example, are described in Ref \cite{tkacik+al_07}. It is worth noting that a parallel discussion of optimization principles for information transmission has a long history in the context of neural coding, where we can think of the distribution on inputs as given by the sensory environment and optimization is used to predict the form of the input/output relation \cite{barlow_61,laughlin_81,atick+redlich_90,brenner_00}. Although there are many open questions, it would be attractive if a single principle could unify our understanding of information flow across such a wide range of biological systems. 

\begin{acknowledgments}
{\small We thank T Gregor, JB Kinney, DW Tank and EF Wieschaus for helpful discussions. This work was supported in part by NIH grants P50 GM071508 and R01 GM077599, NSF grant PHY-0650617, by the Swartz Foundation, the Burroughs Wellcome Fund Program in Biological Dynamics (GT) and  by US Department of Energy grant DE-FG02-91ER40671 (CC).}
\end{acknowledgments}
\appendix{}
\section{Finding optimal channel capacities}
\label{ap:opt}
%
%

If we treat the kernel on a discrete $(c,g)$ grid we can easily choose such $p(c)$ as to maximize the mutual information $\mathrm{I}(c;g)$ between the expression level and the concentration. The problem can be stated in terms of the following variational principle:
\begin{equation}
\mathcal{L}[p(c)]=\sum_{c,g}p(g|c)p(c)\log_2\frac{p(g|c)}{p(g)} - \Lambda\sum_c p(c), \label{func}
\end{equation}
where the multiplier $\Lambda$ enforces the normalization of $p(c)$, and $p(g)$ itself is a function of the unknown distribution (since $p(g)=\sum_c p(g|c)p(c)$). The solution $p^*(c)$ of this problem achieves the capacity, $\mathrm{I}(c;g)$, of the channel.

The original idea behind the Blahut-Arimoto approach \cite{blahut} was to understand that the maximization of Eq (\ref{func}) using variational objects $p(c_i)$ is equivalent to the following maximization:
\begin{equation}
\max_{p(c)} \mathcal{L}[p(c)] \sim \max_{p(c)}\max_{p(c|g)} \mathcal{L}'[p(c),p(c|g)],
\end{equation}
where
\begin{equation}
\mathcal{L}'[p(c),p(c|g)]=\sum_{g,c}p(c)p(g|c)\log\frac{p(c|g)}{p(c)}-\Lambda\sum_c p(c).
\label{baa1}
\end{equation}
In words, finding the extremum in variational object $p(c)$ is equivalent to a double maximization of a modified Lagrangian, where both $p(c)$ and $p(c|g)$ are treated as independent variational objects.
The extremum of the modified Lagrangian is achieved exactly when the consistency condition $p(c|g)=\frac{p(g|c)p(c)}{\sum_c p(g|c)p(c)}$ holds. This allows us to make an iterative algorithm that we detail below, where Eq (\ref{baa1}) is solved for the optimal $p(c)$ and evaluated at some ``known'' $p(c|g)$, which is in turn updated with the newly obtained estimate of $p(c)$.

Before describing the algorithm let us also suppose that each input signal $c$ carries some metabolic or time cost to the cell. Then we can introduce a cost vector $v(c)$ that assigns a cost to each codeword $c$, and require of the solution the following:
\begin{equation}
\sum_c p(c)v(c)\leq C_0,
\end{equation}
where $C_0$ is the maximum allowed expense. The constraint can be introduced into the functional,  Eq (\ref{func}) or Eq (\ref{baa1}), through an appropriate Lagrange multiplier; the same approach can be taken to introduce the cost of coding for the output words, $\sum_g \sum_c p(g|c) p(c) v(g)$, because it reduces to an additional ``effective'' cost for the input, $\tilde{v}(c)=\sum_g p(g|c) v(g)$. 

As was pointed out in the main text, after discretization we have no guarantees that the optimal distribution $p(c_i)$ is going to be smooth.
One way to address this problem is to enforce the smoothness on the scale set by the precision at which the input concentration can be controlled by the cell, $\sigma_c(\bar{c})$, by penalizing big derivatives in the Lagrangian of Eq (\ref{baa1}). An alternative way is to find the spiky solution (without imposing any direct penalty term), but interpret it not as a real, ``physical'' concentration distribution, but rather as the distribution of concentrations that the cell attempts to generate, $c^*$. In this case, however, the limited resolution of the input, $\sigma_c(\bar{c})$, must be referred to the output as an additional effective noise in gene expression, $\sigma_g^2=\sigma_c^2 \left(\frac{\partial \bar{g}}{\partial c}\right)^2$. The optimal solution $p(c^*)$ is therefore the distribution of the levels that the cell would use if it had infinitely precise control over choosing various $c^*$ (i.e. if the input noise were absent), but the physical concentrations are obtained by convolving this optimal result $p(c^*)$ with a Gaussian of width $\sigma_c(c^*)$. Although we chose to use the second approach to compute the results of this paper, we will, for completeness, describe next how to include the smoothness constraint into the functional explicitly.

If the smoothness of the input distribution $p(c)$ is explicitly constrained in the optimization problem, then it will be controllable through an additional Lagrange multiplier, and both ways of computing the capacity  -- that of referring the limited input resolution $\sigma_c(\bar{c})$ to the noise in the output, and that of including it as a smoothness constraint on the input distribution -- will be possible within a single framework.
We proceed by analogy to field theories in which the kinetic energy terms of the form $\int |\nabla f(x)|^2\,dx$ constrain the gradient magnitude, and form the following functional:
\begin{eqnarray}
\mathcal{L}[p(c)]&=&\mathrm{I}(c;g) - \lambda_0 \sum_c p(c) - \label{sba1}\\
&-&\Phi_1 \sum_c p(c) v_1(c)-\Phi_2 \sum_g p(g)v_2(g) - \label{sba2}\\
&-& \Theta \sum_c \left(\frac{\Delta p}{\Delta c}\sigma(c)\right)^2 \label{sba3}.
\end{eqnarray}
Eq (\ref{sba1}) maximizes the capacity with respect to variational objects $p(c)$ while keeping the distribution normalized; Eq (\ref{sba2}) imposes cost $v_1(c)$ on input symbols and cost $v_2(g)$ on output symbols; finally, Eq (\ref{sba3}) limits the derivative of the resulting solution. The difference operator $\Delta$ is defined for an arbitrary function $f(c)$:
\begin{equation}
\Delta f(c)=f(c_{i+1})-f(c_i).
\end{equation}
$\sigma(c)$ assigns a different weight to various intervals on the input axis, $c$.  If the input cannot be precisely controlled, but has an uncertainty of $\sigma(c)$ at mean input level $c$, we require that the optimal probability distribution must not change much as the input fluctuates on the scale $\sigma(c)$. In other words, we require for each input concentration that:
\begin{equation}
\delta p = \frac{\Delta p}{\Delta c}\sigma(c) \ll 1;
\end{equation}
the term in Eq.\ref{sba3} constrained by Lagrange multiplier $\Theta$ can be seen as the sum of squares of such variations over all possible values of the input.
\begin{widetext}
By differentiating the functional, Eq (\ref{baa1}), that includes the relevant constraints, with respect to $p(c_i)$ we get the following equation:
\begin{eqnarray}
0&=&\sum_g p(g|c_i)\log p(c_i|g) - \log p(c_i) - \lambda - \Phi_1 v_1(c_i) - \Phi_2 \sum_g p(g|c_i)v_2(g) + \\
&+&\Theta \left\{\left[p(c_{i+1})-p(c_i)\right]\frac{\sigma^2(c_i)}{(c_{i+1}-c_i)^2}-\left[p(c_{i})-p(c_{i-1})\right]\frac{\sigma^2(c_{i-1})}{(c_{i}-c_{i-1})^2}\right\}.
\end{eqnarray}
Let us denote by $F(c,p(c))=\Delta \frac{\Delta p}{(\Delta c)^2}\sigma^2$ the term in braces. The solution for $p(c)$ is therefore given by:
\begin{eqnarray}
p(c)&=&\frac{1}{Z}\exp
\left\{
\sum_g p(g|c)\log p(c|g) -\Phi_1 v_1(c) - \Phi_2\sum_gp(g|c)v_2(g) +\Theta F(c,p(c))
\right\}. \label{sba4}
\end{eqnarray} 
We can now continue to use the Blahut-Arimoto trick of pretending that $p(c|g)$ is an independent variational object, and that $p(c)$ has to be solved with $p(c|g)$ held fixed; however, even in that case, Eq (\ref{sba4}) is an implicit equation for $p(c)$ which needs to be solved by numerical means. The complete iterative prescription is therefore as follows:
\begin{eqnarray}
p^{n}(g)&=&\sum_c p(g|c)p^{n}(c) \\
p^n(c|g)&=&\frac{p(g|c) p^n(c)}{p^n(g)}\\
p^{n+1}(c)&=&\frac{1}{Z}\exp
\left\{
\sum_g p(g|c)\log p^n(c|g) -\Phi_1 v_1(c) - \Phi_2\sum_gp(g|c)v_2(g) +\Theta F(c,p^{n+1}(c))
\right\}. \label{iter1}
\end{eqnarray}
\end{widetext}

Again, Eq (\ref{iter1}) has to be solved on its own by numerical means as the variational objects for iteration $(n+1)$ appear both on the left- and right-hand sides. The input and output costs of coding are neglected if one sets $\Phi_1=\Phi_2=0$; likewise, smoothness constraint is ignored for $\Theta=0$, in which case Eq (\ref{iter1}) is the same as in the original Blahut-Arimoto derivation and it gives the value of $p^{n+1}(c)$ explicitly.

For the capacities computed in this paper we have calculated the effective output noise that includes the intrinsic output noise as well as the input noise that has been referred to the output (see Section \ref{sandn}); we can therefore set $\Theta=0$. This approach treats all sources of noise on the same footing and allows us to directly compare the magnitudes of noise sources at the input and the output.
We also note that  it makes sense to compute and compare the optimal distribution of outputs rather than inputs: the input/output kernels are degenerate and there are various input distributions (differing either in the regions that give saturated or zero response, or by having variations on a scale below $\sigma_c$) that will yield essentially the same distribution of outputs.

\section{Validity of Langevin approximations}
\label{ax:lan}
Langevin approximation assumes that the fluctuations of a quantity around its mean are Gaussian and proceeds to calculate their variance \cite{vanKampen_07}. For the calculation of exact channel capacity we must calculate the full input/output relation, $p(g|c)$. Even if Langevin approach ends up giving the correct variance as the function of the input, $\sigma_g(c)$, the shape of the distribution might be far from Gaussian. We expect such a failure when the number of mRNA is very small: the distribution of expression levels might be then multi-peaked, with peaks corresponding to $b,2b,3b,\dots$ proteins, where $b$ is the burst size (number of proteins produced per transcript lifetime). 

In the model used in Eq (\ref{actnoise}), parameter $\alpha=(1+b)/g_0$ determines the output noise; $g_0=b \bar{m}$, where $\bar{m}$ is the average number of transcripts produced during the integrating time (i.e. the longest averaging timescale in the problem, for example the protein lifetime or cell doubling time). If $b\gg 1$, then the output noise is effectively determined only by the number of transcripts, $\alpha\approx 1/\bar{m}$. We should therefore be particularly concerned what happens as $\bar{m}$ gets small. 

Our plan is therefore to solve for $p(g|c)$ exactly by finding the stationary solution of the Master equation in the case where the noise consists of the output and switching input contributions. In this approach, we explicitly treat the fact that the number of transcribed messages, designated by $m$, is discrete. We start by calculating $p_i(m|c,t)$. The state of the promoter is described index  $i$, which can be 0 or 1, depending on whether the promoter is bound by the transcription factor or not, respectively. Normalization requires that for each value of $c$:
\begin{equation}
\sum_{i=0,1}\sum_mp_i(m|c,t)=1.
\end{equation}
The time evolution of the system is described by the following set of equations for an activator:
\begin{eqnarray}
\frac{\partial p_0(m|c,t)}{\partial t}&=&R_e\left(p_0(m-1|c,t)-p_0(m|c,t)\right)	\nonumber \\	
&-&\frac{1}{\tau}\left(m p_0(m|c,t)-(m+1)p_0(m+1|c,t)\right)	\nonumber \\
&-&k_- p_0(m|c,t) + k_+ c p_1(m|c,t),				 \\
\frac{\partial p_1(m|c,t)}{\partial t}&=&-\frac{1}{\tau}\left(m p_1(m|c,t)-(m+1)p_1(m+1|c,t)\right) \nonumber	\\
&+&k_- p_0(m|c,t) - k_+ c p_1(m|c,t),
\end{eqnarray}
where $\tau$ is the integrating time, $k_-$ is the rate for switching into the inactive state (off-rate of the activator), $k_+$ is the second-order on-rate, and $R_e$ is the rate of mRNA synthesis. These constants combine to give $\bar{m}=R_e\tau$ and the input switching noise strength $\gamma=(k_-\tau)^{-1}$, see Table \ref{t-params}.
This set of equations is supplemented by appropriate boundary conditions for $m=0$. To find a steady state distribution $p(m|c, t\rightarrow \infty)=p(m|c)$, we set the left-hand side to zero and rewrite the set of equations (with high enough cutoff value of $m_\mathrm{max}$) in matrix form:
\begin{equation}
{\mathbf M}(c) {\mathbf p}(c)={\mathbf b},
\end{equation}
where ${\mathbf p}=\left(p_0(0|c), p_1(0|c), p_0(1|c), p_1(1|c), \cdots\right)$ and ${\mathbf b}=(0,0,\cdots, 0, 1)$. Matrix ${\mathbf M}$ (of dimension $2(m_\mathrm{max}+1) + 1$ rows and $2(m_\mathrm{max}+1)$ columns) contains, in its last row, only ones, which enforces normalization. The resulting system is a non-singular band-diagonal system that can be easily inverted. The input/output relation for the number of messages is given by taking $p(m|c)=p_0(m|c)+p_1(m|c)$.

Having found the distribution for the number of transcripts we then convolve it another Poisson process, $p(g|\langle g \rangle = b\,m)$, i.e. $p(g|c)=\sum_m p(m|c) p(g| \langle g \rangle = b\,m)$. Finally, the result is rediscretized such that mean expression $\bar{g}$ runs from 0 to 1.

Note that the Langevin approximation only depends on the combination of the burst size $b$ and the mean number of transcripts $\bar{m}$ through $\alpha$; in contrast, the Master equation solution depends on both $b$ and $\bar{m}$ independently.
The generalization of this calculation to repressors or Hill-coefficient-type cooperativity is straightforward.

\begin{figure} 
   \centering
   \includegraphics[width=3.3in]{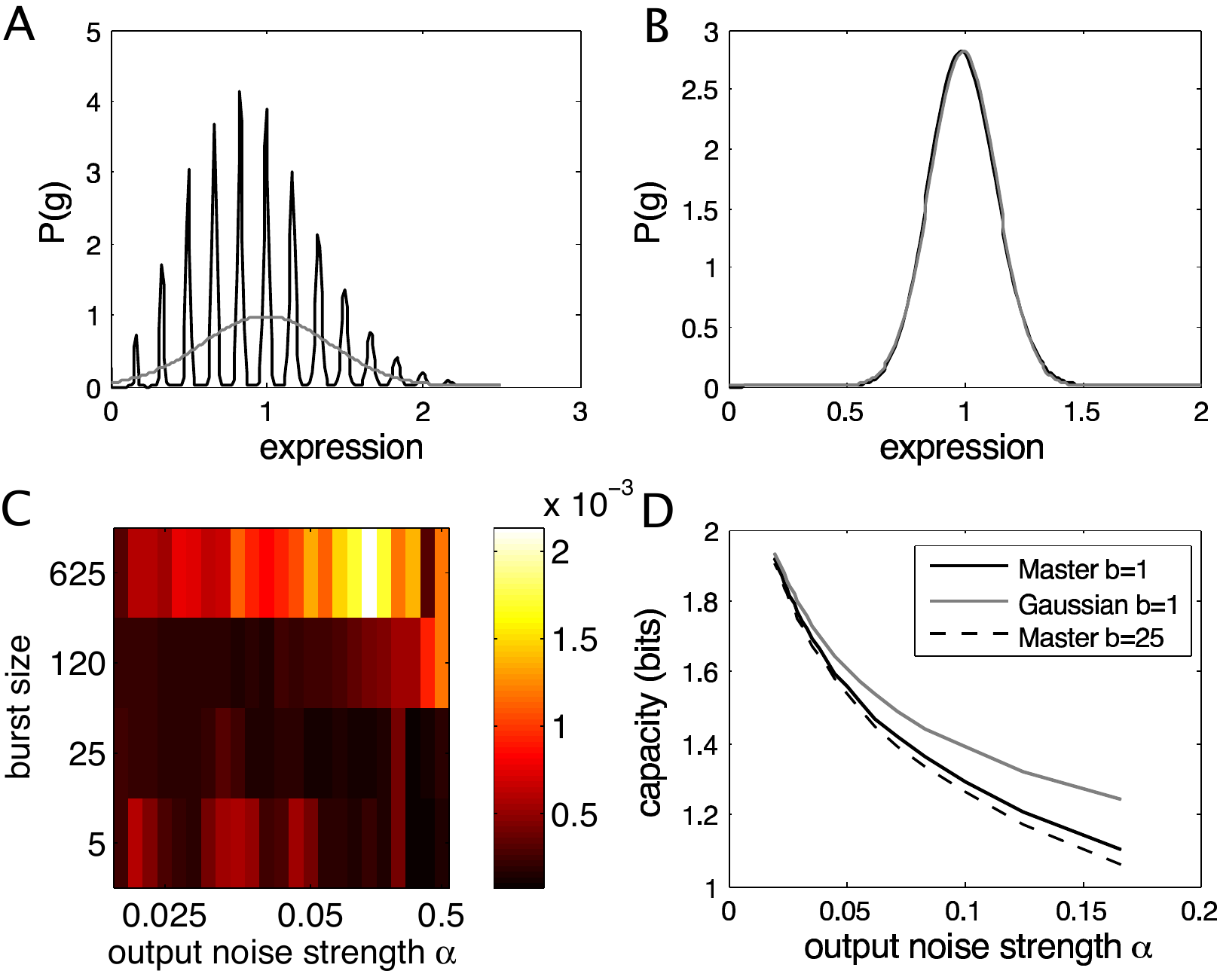} 
   \caption{(Color online) Exact solutions (black) for input/output relations, $p(g|c)$, compared to their Gaussian approximations (gray). Panel A shows the distribution of outputs at maximal induction, $p(g|c_{\rm max})$ for a system with a large burst size, $b=5^4$ and a large output noise $\alpha=\frac{1}{6}$ (i.e. the average number of messages is ~6, as is evident from the number of peaks, each of which corresponds to a burst of translation at different number of messages). Panel B shows the same distribution for smaller output noise, $b=5^2$ and $\alpha=\frac{1}{50}$; here Gaussian approximation performs well. Both cases are computed with switching noise parameter $\gamma=\frac{1}{50}$, and cooperativity of $h=2$. Panel C shows in color-code the error made in computing the standard deviation of the output given $c$; the error measure we use is the maximum difference between the exact and Gaussian results over the full range of concentrations: ${\rm max}_c \, {\rm abs}\left\{[\sigma_g(c)/g_0]_{\rm Master}-[\sigma_g(c)/g_0]_{\rm Gaussian}\right\}$. As expected the error decreases with decreasing output noise. Panel D shows that the capacity is overestimated by using an approximate kernel, but the error again decreases with decreasing noise as Langevin becomes an increasingly good approximation to the true distribution. In the worst case the approximation is about $12\%$ off. Gaussian computation only depends on $\alpha$ and not separately on burst size, so we plot only one curve for $b=1$.}
   \label{f-master}
   \end{figure}

Fig \ref{f-master}c shows that the Langevin approximation yields correct second moments of the output distribution; however, Gaussian distributions themselves are, for large burst sizes and small number of messages, inconsistent with the exact solutions, as can be seen in Fig \ref{f-master}a. In the opposite limit where the number of messages is increased and burst size kept small, see Fig \ref{f-master}b, normal distributions are an excellent approximation. Despite these difficulties the information capacity calculated with either Gaussian or Master input/output relations differs by at most $12\%$ over a large range of burst sizes $b$ and values for $\alpha$, illustrated by Fig \ref{f-master}d.
\section{Fine-tuning of optimal distributions}
\label{ax:pert}
\begin{figure}[!h] 
   \centering
   \includegraphics[width=3.5in]{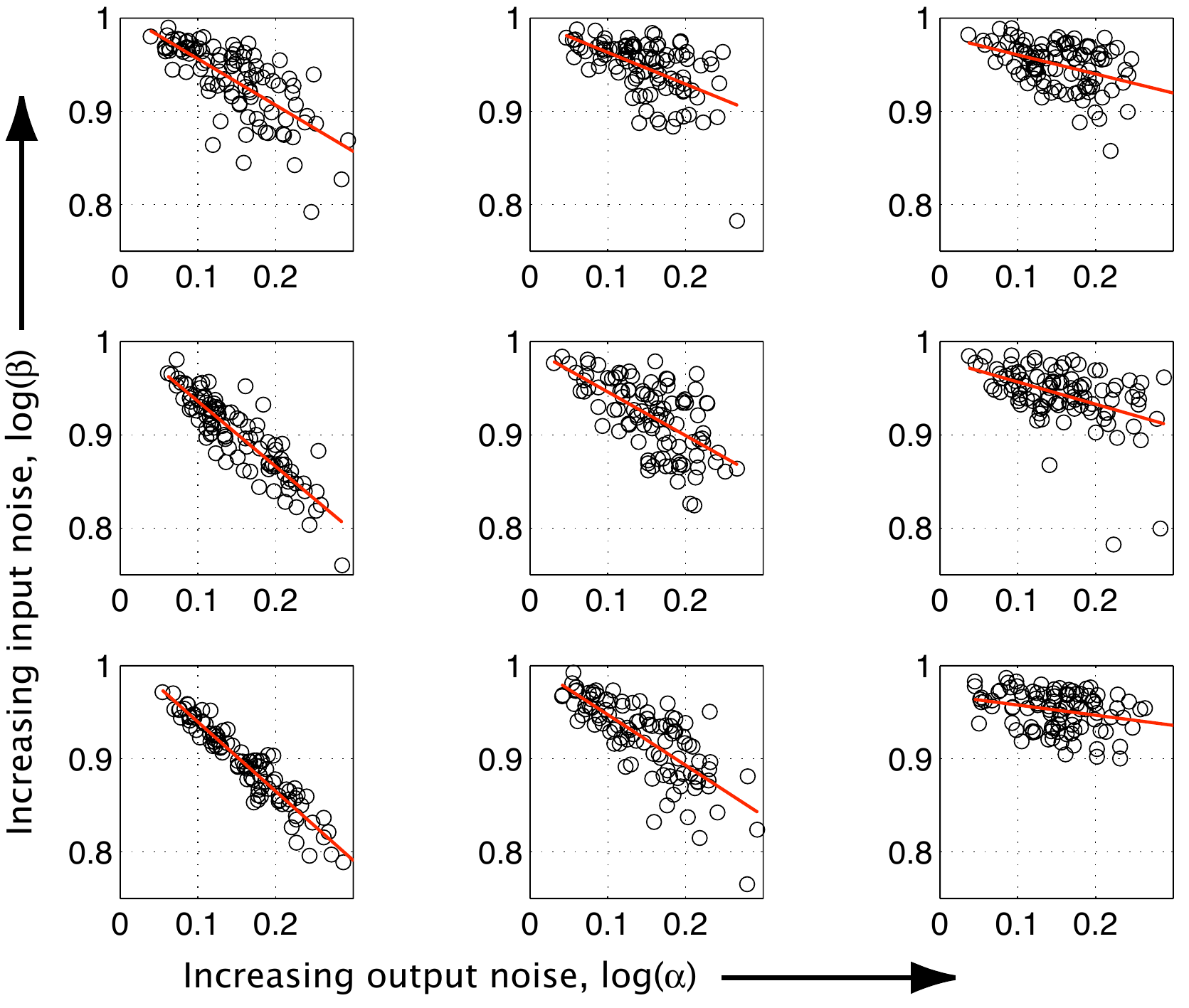} 
   \caption[(Color online) Robustness of the optimal solutions.]{Robustness of the optimal solutions to perturbations in the input distribution. Activator systems with no cooperativity are plotted; their parameters are taken from an uniformly spaced, $3\times 3$ grid of points in the noise plane of Fig \ref{f-cpanel1}a, such that the output noise increases along the horizontal edge of the figure and the input noise along the vertical edge. Each subplot shows a scatter plot of 100 perturbations from the ideal solution; the Jensen-Shannon distance from the optimal solution, $d_i$, is plotted on the horizontal axis and the channel capacity (normalized to maximum when there is no perturbation), $I_i/I_{\rm max}$, on the vertical axis. Red lines are best linear fits.}
   \label{f-optperturb}
   \end{figure}

To examine the sensitivity to the perturbations in the optimal input distributions for Fig \ref{f-costs_x} we need to generate an ensemble of perturbations. We pick an \emph{ad hoc} prescription, whereby the optimal solution is taken, and we add to it 5 lowest harmonic modes  on the input domain, each with a weight that is uniformly distributed on some range. The range determines whether the perturbation is small or not. The resulting distribution is clipped to be positive and renormalized. This choice was made to induce low-frequency perturbations (high frequency perturbations get averaged out because the kernel is smooth). Then, for an ensemble of 100 such perturbations, $p_i(c),\,\, i=1,\dots,100$, and for every system of the noise plane in Fig \ref{f-cpanel1}a, the divergence of the perturbed input distribution to the true solution, $d_i=D_{\rm JS}(p_i(c),p^*(c))$, is computed, as well as the information transmission, $I_i=I[p(g|c),p_i(c)]$. Figure \ref{f-optperturb} plots the $(d_i,I_i)$ scatter plots for $3\times 3$ representative systems with varying amounts of output ($\alpha$) and input ($\beta$) noise, taken from Fig \ref{f-cpanel1}a uniformly along the horizontal and vertical axes. 

 Figure \ref{f-optperturb} shows that as we move towards systems with higher capacity (lower left corner), perturbations to the optimal solution that are at the same distance from the optimum as in the low capacity systems (upper right corner), will cause greater relative loss (and therefore an even greater absolute loss) in capacity. As expected, higher capacity systems must be better tuned, but even for the highest capacity system considered, a perturbation of around $d_{\rm JS}\approx 0.2$ will only cause an average $15\%$ loss in capacity. We also note that for systems with high capacity the linear relationship between the the divergence $d_i$ and capacity $I_i$ provides a better fit than for systems with small capacity.
\section{Nonspecific binding}
\label{ax:nb}
One needs to make a careful distinction between the total concentration of the input transcription factors, $c_t$, and the free concentration $c_f$, diffusing in solution in the nucleus. We imagine the true binding site embedded in a pool of non-specific binding sites -- perhaps all other short fragments of DNA -- and there being an ongoing competition between one functional site (with strong affinity) and large number of weaker non-specific sites. If these non-specific sites are present at concentration $\rho$ in the cell, and have affinities drawn from some distribution $p(K)$, the relationship between the free and the total concentration of the input is:
\begin{equation}
c_t = c_f + \rho \int dK \, p(K)\, \frac{c_f}{c_f+K}. \label{nscost}
\end{equation}
Importantly, the concentration that enters all information capacity calculations is the 
\emph{free} concentration $c_f$, because it directly determines both the promoter occupancy in Eq (\ref{meanio}) as well as diffusive noise; on the other hand, the cell can influence the free concentration only by producing more or less of the transcrption factor, i.e. by varying (and paying for) the \emph{total} concentration. If the free concentration is well below the strength of the non-specific binding, $\langle K\rangle$, Eq (\ref{nscost}) can be approximated by $c_t\approx c_f (1+ \rho/\langle K\rangle)$, with the total and free concentrations being proportional to each other. Because the cost functional, Eq (\ref{cconstraint}), is only determined to within a factor anyway, the presence of non-specific sites will effectively just rescale the cost per free molecule of transcription factor. A separate calculation is needed to show that the presence of non-specific binding does not appreciably increase the noise in gene expression (to be presented elsewhere).

\end{document}